\documentclass[prx,floatfix,showpacs,superscriptaddress, twocolumn]{revtex4-2} 
\usepackage{booktabs}
\usepackage{graphicx}
\usepackage{dcolumn}
\usepackage{bm}
\usepackage{multirow}
\usepackage{xcolor}
\usepackage{amsbsy}
\usepackage{amsmath}  
\usepackage{tikz}
\newcommand{\bibcommenthead}[1]{}
\usepackage[colorlinks=true,
linkcolor=blue,
anchorcolor=blue,
urlcolor=blue,
citecolor=blue]{hyperref} 
\usepackage{amssymb}

\begin{document}

\title{Skyrmion and Meron Crystals in Intermetallic Gd$_3$Ru$_4$Al$_{12}$: Microscopic Model Insights into Chiral Phases}
\author{Jiajun Mo}
\affiliation{Department of Physics, University of Science and Technology of China, Hefei, Anhui 230026, People's Republic of China}

\author{Leandro M. Chinellato}
\author{Fletcher Williams}
\affiliation{Department of Physics and Astronomy, University of Tennessee, Knoxville, TN 37996, USA}
\affiliation{Materials Science \& Technology Division, Oak Ridge National Laboratory, Oak Ridge, TN 37831, USA}

\author{Akiko Kikkawa}
\affiliation{RIKEN Center for Emergent Matter Science (CEMS), Wako, 351-0198, Japan}

\author{Joseph A. M. Paddison}
\affiliation{Materials Science \& Technology Division, Oak Ridge National Laboratory, Oak Ridge, TN 37831, USA}

\author{Matthias D. Frontzek}
\affiliation{Neutron Scattering Division, Oak Ridge National Laboratory, Oak Ridge, TN 37831, USA}

\author{Gabriele Sala}
\affiliation{Neutron Scattering Division, Oak Ridge National Laboratory, Oak Ridge, TN 37831, USA}

\author{Chris Pasco}
\affiliation{Materials Science \& Technology Division, Oak Ridge National Laboratory, Oak Ridge, TN 37831, USA}

\author{Kipton Barros}
\affiliation{Theoretical Division, Los Alamos National Laboratory, Los Alamos, NM 87545, USA}

\author{Taro Nakajima}
\affiliation{RIKEN Center for Emergent Matter Science (CEMS), Wako, 351-0198, Japan}
\affiliation{Institute for Solid State Physics, The University of Tokyo, Chiba 277-8581, Japan}

\author{Taka-hisa Arima}
\affiliation{RIKEN Center for Emergent Matter Science (CEMS), Wako, 351-0198, Japan}
\affiliation{Department of Applied Physics and Quantum-Phase Electronics Center (QPEC), The University of  Tokyo, Bunkyo-ku, Tokyo 113-8656, Japan}
\affiliation{Department of Advanced Materials Science, University of Tokyo, Japan}

\author{Yasujiro Taguchi}
\affiliation{RIKEN Center for Emergent Matter Science (CEMS), Wako, 351-0198, Japan}

\author{Yoshinori Tokura}
\affiliation{RIKEN Center for Emergent Matter Science (CEMS), Wako, 351-0198, Japan}
\affiliation{Department of Applied Physics, University of Tokyo, Tokyo 113-8656, Japan}
\affiliation{Tokyo College, University of Tokyo, Tokyo 113-8656, Japan}

\author{Matthew B. Stone}
\affiliation{Neutron Scattering Division, Oak Ridge National Laboratory, Oak Ridge, TN 37831, USA}

\author{Andrew D. Christianson}
\affiliation{Materials Science \& Technology Division, Oak Ridge National Laboratory, Oak Ridge, TN 37831, USA}

\author{Cristian D. Batista}
\affiliation{Department of Physics and Astronomy, University of Tennessee, Knoxville, TN 37996, USA}
\affiliation{Neutron Scattering Division, Oak Ridge National Laboratory, Oak Ridge, TN 37831, USA}

\author{Shang Gao}%
\email{sgao@ustc.edu.cn}
\affiliation{Department of Physics, University of Science and Technology of China, Hefei, Anhui 230026, People's Republic of China}
\affiliation{Materials Science \& Technology Division, Oak Ridge National Laboratory, Oak Ridge, TN 37831, USA}
\affiliation{Neutron Scattering Division, Oak Ridge National Laboratory, Oak Ridge, TN 37831, USA}

\date{\today}
              
\begin{abstract} 
Topological spin textures in frustrated intermetallics hold great promise for spintronics applications. However, understanding their origin and properties remains a significant challenge due to competing and often long-range interactions mediated by conduction electrons. Here, by combining neutron scattering experiments with theoretical modeling via unprecedented multi-target fits that further incorporate the ferromagnentic resonance data and magnetization curve, we construct a realistic microscopic model for the prototypical intermetallic skyrmion host \text{Gd}$_3$\text{Ru}$_4$\text{Al}$_{12}$. Beyond magnetic frustration, we identify the competition between dipolar interactions and easy-plane single-ion anisotropy as a key ingredient for stabilizing the rich chiral magnetic phases observed in this compound---including a hexagonal skyrmion crystal and two distinct meron crystals. Remarkably, the meron crystal in lower field is revealed to be commensurate with the underlying lattice, and its unique three-meron-one-antimeron spin texture is verified by the polarized x-ray diffraction data. At elevated temperatures, the short-range spin correlations in \text{Gd}$_3$\text{Ru}$_4$\text{Al}$_{12}$ are well described by a codimension-two spiral spin-liquid. Perturbations from staggered Dzyaloshinskii-Moriya interactions give rise to chiral fluctuations that account for the temperature and field dependence of the anomalous Hall response. Our results highlight the unique power of neutron scattering, especially when combined with complementary experimental techniques, to unravel complex magnetic phase transitions and provide new insights into the rich variety of topological spin textures in frustrated systems.
\end{abstract} 

\maketitle
 
\section{\label{sec:level1} Introduction}
 
The discovery of magnetic skyrmions in the non-centrosymmetric compound MnSi~\cite{muhlbauerSkyrmion2009} has sparked broad interest in developing novel spintronics devices based on topological spin textures \cite{fertMagnetic2017, zhang_skyrmion_2020, reichhardt_statics_2022}. Their nanoscale size and topologically protected stability make magnetic skyrmions particularly attractive as a medium for data storage \cite{nagaosaTopological2013, back_skyrmionics_2020}.
As demonstrated in various prototype devices, intermetallic compounds offer distinct advantages for controlling skyrmions \cite{sampaioNucleation2013, zhang_skyrmion_2020, chen_all_2024}. Information encoded in these topological spin structures can be written, manipulated, and erased simply by applying an electric current \cite{schulz2012emergent, yuCurrentInduced2017, zhang_skyrmion_2020}.
In addition, the intrinsic chirality of skyrmions couples to conduction electrons, generating  emergent magnetic fields that can reach hundreds of Tesla \cite{nagaosaTopological2013, kurumajiSkyrmion2019}. This leads to pronounced anomalous electronic and thermal transport responses, which are crucial for future electronic applications \cite{nagaosa_anomalous_2010, kimbell2022challenges, wang_topological_2022}.

From a fundamental perspective, the interaction between localized moments and conduction electrons introduces additional challenges in understanding the origin and properties of topological spin textures~\cite{hayami_topological_2021}.  Modeling such systems in magnetic metals is particularly difficult due to the presence of long-range interactions mediated by conduction electrons, such as the Ruderman–Kittel–Kasuya–Yosida (RKKY) interaction~\cite{ruderman_indirect_1954,kasuya_1956, yosida_magnetic_1957}. When the wavelength of the spin texture is much longer than the lattice space,  some of these challenges can be alleviated using effective field theories~\cite{rosslerSpontaneous2006,yu2010real}. However, this approach breaks down in many recently discovered centrosymmetric materials~\cite{kurumajiSkyrmion2019, hirschbergerSkyrmion2019, khanhNanometric2020, gaoFractional2020, takagi2022square, yoshimochiMultistep2024}, where the spin textures span only a few lattice spacings. In such cases, a microscopic, lattice-resolved description becomes essential to capture the emergence of novel phenomena, including topological charge fractionalization~\cite{gaoFractional2020,wangMeron2021a}, quadrupolar correlations \cite{amari_cp2_2022, zhangCP22023}, and the role of quantum fluctuations \cite{lohani_quantum_2019,HallerQuantum2024}.

\begin{figure*}[t]
    \centering
    \hypertarget{anchor:fig1-top}{}
    \includegraphics[width=1\linewidth]{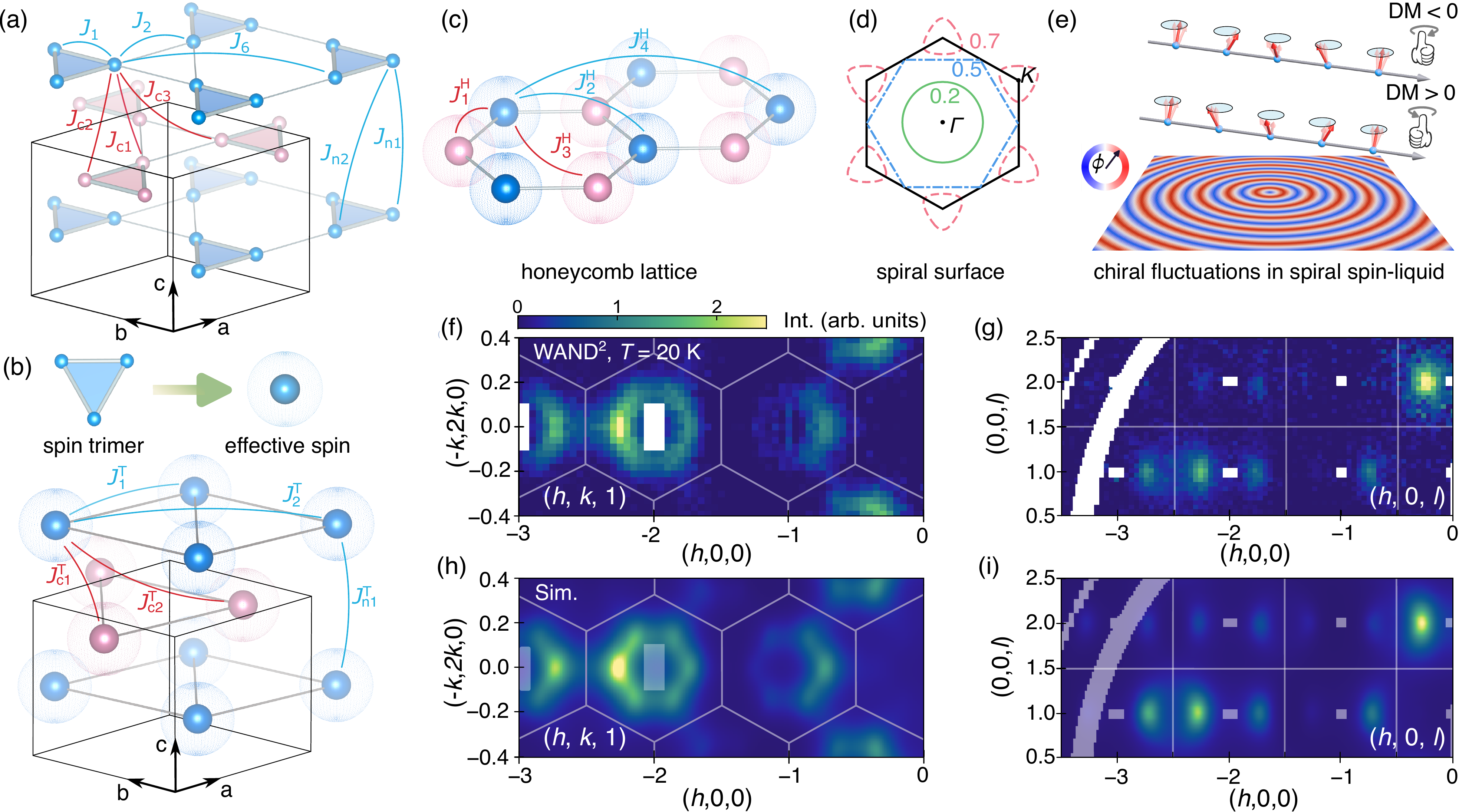}
    \caption{Codimension-two spiral spin-liquid in Gd$_3$Ru$_4$Al$_{12}$. (a) The Gd$^{3+}$ ions with spin \textit{S} = 7/2 in Gd$_3$Ru$_4$Al$_{12}$ form breathing kagome lattices in the $ab$ plane, which are AB-stacked along the $c$ axis. Curved solid lines in the figure indicate the exchange paths for the two shortest intralayer couplings ($J_1$ and $J_2$), the sixth-neighbor intralayer coupling ($J_6$), the three shortest interlayer couplings ($J_{\rm{c1}}$, $J_{\rm{c2}}$, and $J_{\rm{c3}}$), and the two shortest second-layer couplings ($J_{\rm{n1}}$ and $J_{\rm{n2}}$). (b) Under dominant ferromagnetic $J_1$ couplings, the three spins over the smaller triangles of the breathing kagome lattice can be viewed as an effective spin $S_{\rm{eff}}=3S$, leading to AB-stacked triangular lattices of effective spins. Curved solid lines indicate the exchange paths for the two shortest intralayer couplings ($J_1^{\rm{T}}$ and $J_2^{\rm{T}}$), the two shortest interlayer couplings ($J_{\rm{c1}}^{\rm{T}}$ and $J_{\rm{c2}}^{\rm{T}}$), and the shortest second-layer coupling ($J_{\rm{n1}}^{\rm{T}}$). (c) Equivalent honeycomb lattice formed by the effective spins. Curved solid lines indicate the exchange paths for the inter-sublattice couplings ($J_1^{\rm{H}}$ and $J_3^{\rm{H}}$) and the intra-sublattice couplings ($J_2^{\rm{H}}$ and $J_4^{\rm{H}}$). (d) Representative spiral rings in reciprocal space calculated for a $J_1^{\rm{H}}$-$J_2^{\rm{H}}$ honeycomb-lattice model with a frustration ratio of $|J_2^{\rm{H}}/J_1^{\rm{H}}|=0.2$ (green), 0.5 (blue), and 0.7 (red). (e) The panel at the bottom describes a real-space spin configuration surrounding a momentum vortex in the spiral spin-liquid state. Pseudocolor corresponds to the phase $\phi$ of the in-plane spin rotation. In the presence of antisymmetric DM interactions, spin fluctuations will become chiral as described in the panel on the top. (f, g) Diffuse neutron scattering patterns in the ($h$,~$k$,~1) (f) and ($h$, 0, $l$) (g) planes measured on WAND$^2$ at $T = 20$~K. Gray lines mark the Brillouin zone boundaries in reciprocal space. (h, i) Simulated diffuse neutron scattering patterns in the ($h$,~$k$,~1) (h) and ($h$, 0, $l$) (i) planes using the classical Monte Carlo simulations for the fitted $J_{126}$-$J_{\rm{c123}}$-$J_{\rm{n12}}$ model on the original breathing kagome lattice.}
    \label{fig:fig1}
\end{figure*}

One prominent example is the centrosymmetric intermetallic material Gd$_3$Ru$_4$Al$_{12}$, in which the magnetic Gd$^{3+}$ ions form a breathing kagome lattice as shown in Fig.~\hyperlink{anchor:fig1-top}{\ref{fig:fig1}(a)}. Combined resonant x-ray diffraction, transport, and transmission electron microscopy (TEM) experiments on this compound have firmly established the existence of magnetic skyrmions with a short pitch of $\sim2$~nm~\cite{nakamuraSpin2018, matsumuraHelical2019, hirschbergerSkyrmion2019}. Despite the many theoretical proposals, which include frustration from the lattice geometry~\cite{hirschbergerSkyrmion2019, kawamura_frustration_2025} or from RKKY interactions~\cite{wangSkyrmion2020, bouazizFermiSurface2022}, multi-spin interactions~\cite{hirschbergerNanometric2021, hayami_stabilization_2024}, and inter-orbital $d$-$f$ frustration~\cite{nomoto_formation_2020, Takuya2023}, a definite explanation for the emergent skyrmion phase through microscopic modeling is still missing. Moreover, compared to the other Gd-based skyrmion hosts~\cite{kurumajiSkyrmion2019, khanhNanometric2020, yoshimochiMultistep2024}, the existence of an easy-plane single-ion anisotropy (SIA)~\cite{hirschbergerSkyrmion2019, hirschbergerLatticecommensurate2024}, which generally destabilizes skyrmions in centrosymmetric systems~\cite{wangMeron2021a, leonovMultiply2015}, makes Gd$_3$Ru$_4$Al$_{12}$ a unique model compound for studying the effects of competing anisotropic interactions—specifically, dipole-dipole interactions versus easy-plane SIA—on topological spin textures. Besides the ordered phases, more recent transport measurements on Gd$_3$Ru$_4$Al$_{12}$ suggest that nonzero chirality  persists in the correlated paramagnetic phase~\cite{kolincioKagome2023}, further underscoring the need for a microscopic model that goes beyond phenomenological approaches.

\begin{figure*}[t!]
    \centering
    \hypertarget{anchor:fig2-top}{}
    \includegraphics[width=1\linewidth]{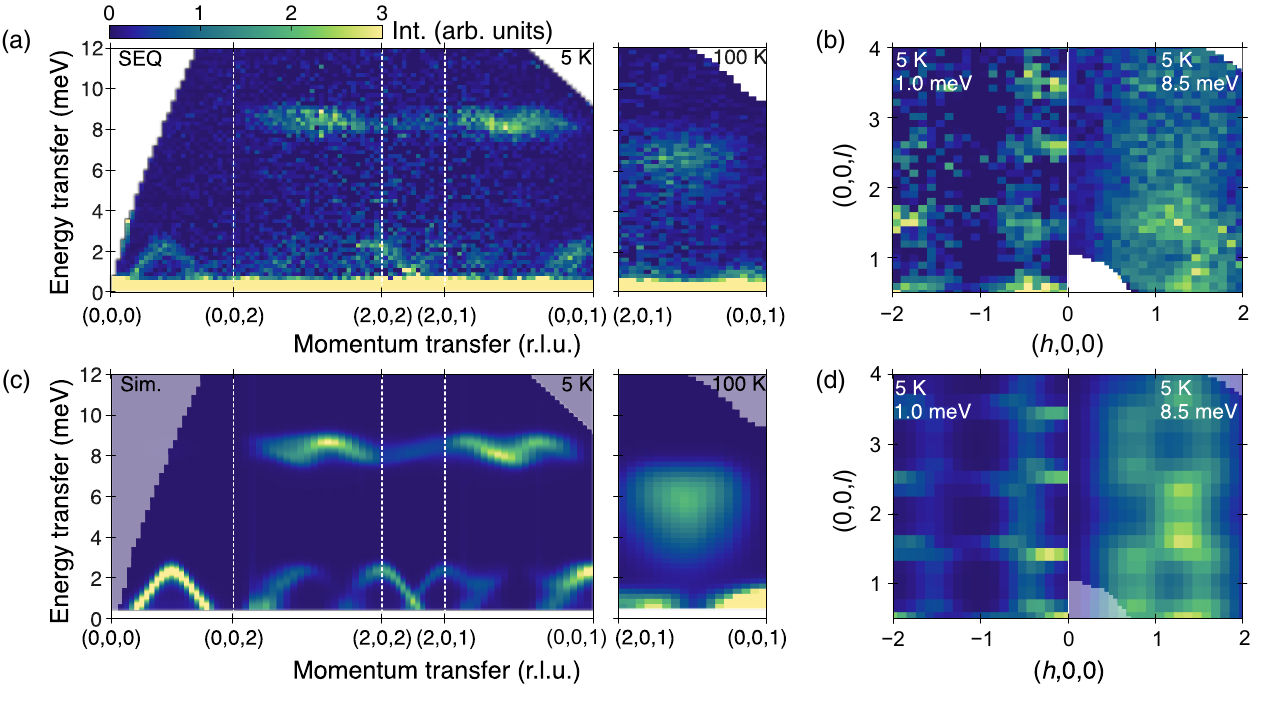}
    \caption{Hierarchical spin excitations in Gd$_3$Ru$_4$Al$_{12}$. (a) INS spectra, $S$(\textbf{Q},~$\omega$), measured on SEQUOIA (SEQ) along the high symmetry lines at $T = 5$~K in the magnetically long-range ordered regime (left) and at $T=100$~K in the correlated paramagnetic regime (right). (b) Experimental constant-energy slices for the INS spectra measured at $T = 5$~K in the ($h$,~0,~$l$) plane at $E$ = 1.0 (left) and 8.5~meV (right). (c) Landau-Lifshitz-Gilbert (LLG) dynamics simulations of the spectra using the fitted $J_{126}$-$J_{\rm{c123}}$-$J_{\rm{n12}}$ model. The simulated spectra have been convoluted by the instrumental energy resolution as described in the Supplemental Material~\cite{supp}. (d) Simulated constant-energy slices in the ($h$,~0,~$l$)  plane at $E$ = 1.0 (left) and 8.5~meV (right). Both the experimental and simulated spectra are integrated over an energy width of $\Delta E = \pm$0.15~meV.}
    \label{fig:fig2}
\end{figure*} 

In this study, we devise a multi-target approach to achieve a fully quantitative, microscopic model for Gd$_3$Ru$_4$Al$_{12}$. While previous attempts to model similar Gd-based skyrmion hosts have struggled to reproduce key experimental features such as the saturation field~\cite{paddisonMagnetic2022}, our multi-target approach, which incorporates not only neutron scattering data performed on isotopically enriched crystals but also the previously reported magnetization~\cite{hirschbergerSkyrmion2019} and ferromagnetic resonance (FMR)~\cite{hirschbergerLatticecommensurate2024} data, achieves unprecedented success in describing all known experimental observations. Our model  accurately captures the key transitions and identifies the anisotropic dipole-dipole interaction as the main stabilization mechanism for the field-induced skyrmion crystal together with a novel inverted meron crystal. Remarkably, our modeling further reveals the existence of a commensurate three-meron-one-antimeron spin texture that is stabilized by the easy-plane SIA~\cite{Lin2015}, which can be validated using the reported polarized x-ray diffraction data~\cite{hirschbergerLatticecommensurate2024}. In the short-range correlated paramagnetic regime, magnetic frustration leads to the emergence of a codimension-two spiral spin-liquid~\cite{bergman_order_2007, yao_generic_2021, gao2024codimensiontwospiralspinliquideffective}, where dominant chiral fluctuations originate from the perturbations of staggered Dzyaloshinskii-Moriya (DM) interactions. The rich variety of chiral phases illustrated in our work shall facilitate the design of spintronics devices based on exotic short- and long-range magnetic orders.

\begin{table*}[t]
    \centering
    \caption{Fitted parameters for the $J_{126}$-$J_{\rm{c123}}$-$J_{\rm{n12}}$ model defined on the original AB-stacked breathing kagome lattice. Also listed are the corresponding parameters on the equivalent AB-stacked triangular lattice and the equivalent honeycomb lattice. The magnetic dipole-dipole interactions are considered through the Ewald summation method. For convenience of comparison, parameters of the equivalent triangular and honeycomb lattice models are scaled by a factor of $(S_\mathrm{eff}/S)^2=9$. Numbers in the parentheses represent the standard uncertainties at the last significant digits.}  
    \label{tab:exchange}
    \begin{tabular}{cccccccc}
    \toprule
    \multicolumn{1}{c}{\bfseries Lattice Type} & 
    \multicolumn{6}{c}{\bfseries Parameters (meV)} \\ 
    \cmidrule(l{3pt}r{3pt}){1-8} 
    \multirow{2}{*}{Original Lattice}& $J_1$ & $J_2$ & $J_6$ & $J_{\mathrm{c1}}$ = $J_{\mathrm{c2}}$ & $J_{\mathrm{c3}}$ & $J_{\mathrm{n1}}$ = $J_{\mathrm{n2}}$ & $K_\mathrm{ab}$\\
    \cmidrule(l{3pt}r{3pt}){2-8}
     & $-0.721(16)$ & $0.033(14)$ & $0.024(1)$ & $-0.057(5)$ & $0.030(12)$ & $0.009(3)$  &0.009(1)\\
    \midrule
    \addlinespace[2mm] 
    
    \multirow{2}{*}{Triangular Lattice} & 
    $\mathrm{--}$ & 
    $J^{\mathrm{T}}_1$\,($J_\mathrm{2}$)& 
    $J^{\mathrm{T}}_2$\,($2J_\mathrm{6}$)& 
    $J^{\mathrm{T}}_{\mathrm{c1}}$\,($4J_\mathrm{c1}$) & 
    $J^{\mathrm{T}}_{\mathrm{c2}}$\,($J_\mathrm{c3}$)& 
    $J^{\mathrm{T}}_{\mathrm{n1}}$\,($9J_\mathrm{n1}$)&
    $K_\mathrm{ab}^\mathrm{T}$\,(3$K_\mathrm{ab}$)\\
    \cmidrule(l{3pt}r{3pt}){2-8} 
    & 
    $\mathrm{--}$ & 
    $0.033(14)$ & 
    $0.048 (3)$ & 
    $-0.228 (21)$ & 
    $0.030 (12)$ & 
    $0.081 (31)$ &
    0.027(3)\\
    
    \midrule
    \addlinespace[2mm]
    
    \multirow{2}{*}{Honeycomb Lattice} & 
    $\mathrm{--}$ &  
    $J^{\mathrm{H}}_2$\,($J^\mathrm{T}_1$)& 
    $J^{\mathrm{H}}_4$\,($J^\mathrm{T}_2$)& 
    $J^{\mathrm{H}}_1$\,($2J^\mathrm{T}_\mathrm{c1}$)&
    $J^{\mathrm{H}}_3$\,($2J^\mathrm{T}_\mathrm{c2}$)& 
    $\mathrm{--}$ &
    $K_\mathrm{ab}^\mathrm{H}$\,(3$K_\mathrm{ab}$)\\
    \cmidrule(l{3pt}r{3pt}){2-8} 
    & 
    $\mathrm{--}$ & 
    $0.033(14)$ & 
    $0.048 (3)$ & 
    $-0.456 (42)$ & 
    $0.060 (24)$ & 
    $\mathrm{--}$  &
    0.027(3)\\

    \bottomrule
    \end{tabular}
    \end{table*}

\section{Lattice geometry and spiral spin-liquids}

As shown in Fig.\hyperlink{anchor:fig1-top}{~\ref{fig:fig1}(a)}, the breathing kagome lattice in Gd$_3$Ru$_4$Al$_{12}$ results in very different couplings among the Gd$^{3+}$ spins. According to  previous magnetic susceptibility and resonant x-ray diffraction experiments \cite{nakamuraSpin2018, matsumuraHelical2019, hirschbergerSkyrmion2019}, spins over the smaller triangles form ferromagnetic (FM) trimers, and the relatively weaker inter-trimer couplings  lead to  antiferromagnetic (AFM) ordering below the N{\'e}el temperature $T_\mathrm{N} \sim 18$~K. An effective description of the low-energy spin dynamics in terms of an AB-stacked triangular lattice of spin trimers (see Fig.\hyperlink{anchor:fig1-top}{~\ref{fig:fig1}(b)}) is obtained by integrating out the intra-trimer excitations. This lattice geometry is known to host codimension-two spiral spin-liquids~\cite{yao_generic_2021, gao2024codimensiontwospiralspinliquideffective}, which can emerge from the competition between inter- and intra-sublattice interactions \cite{bergman_order_2007, gaoSpiral2022,gaoSpiral2017}.

Figures~\hyperlink{anchor:fig1-top}{\ref{fig:fig1}(f,g)} show the diffuse neutron scattering patterns measured on a $^{160}$Gd isotope-enriched single crystal at $T = 20$~K, slightly above  $T_\mathrm{N}$. Details for sample preparation and experiments can be found in the Methods section. The scattering intensity is concentrated in the integer-$l$ planes and forms circular rings centered at the Brillouin zone centers in the ($h$, $k$, 1) plane. This ring-like structure closely resembles the patterns observed in honeycomb lattices~\cite{Soichiro_Okumura_2010,gaoSpiral2022}---prototypical hosts of spiral spin-liquids, as depicted in Figs.~\hyperlink{anchor:fig1-top}{\ref{fig:fig1}(c,d)}. These rings are indicative of a classical spiral spin-liquid state persisting above $T_\mathrm{N}$, with an estimated frustration ratio of approximately 0.2.
As illustrated in Fig.~\hyperlink{anchor:fig1-top}{\ref{fig:fig1}(e)}, on top of the ripple-like spiral correlations in real space~\cite{T_Shimokawa_2019}, nonzero vector chirality, $\langle \mathbf{S}_i \times \mathbf{S}_j \rangle \neq \mathbf{0}$, can be induced by the presence of DM interactions. Consequently, the field-induced magnetization, $\langle \mathbf{M} \rangle \neq \mathbf{0} $, leads  to nonzero scalar spin chirality, $\langle (\mathbf{S}_i \times \mathbf{S}_j)\cdot \mathbf{M}_k \rangle \neq  0 $, on the triangles defined by the sites $ijk$. Elucidating the connection between this scalar spin chirality and the experimentally observed anomalous Hall effect~ \cite{kolincioKagome2023} underscores the need for a more detailed microscopic understanding.

\begin{figure*}[t]
    \centering
    \hypertarget{anchor:fig4-top}{}
    \includegraphics[width=0.9\linewidth]{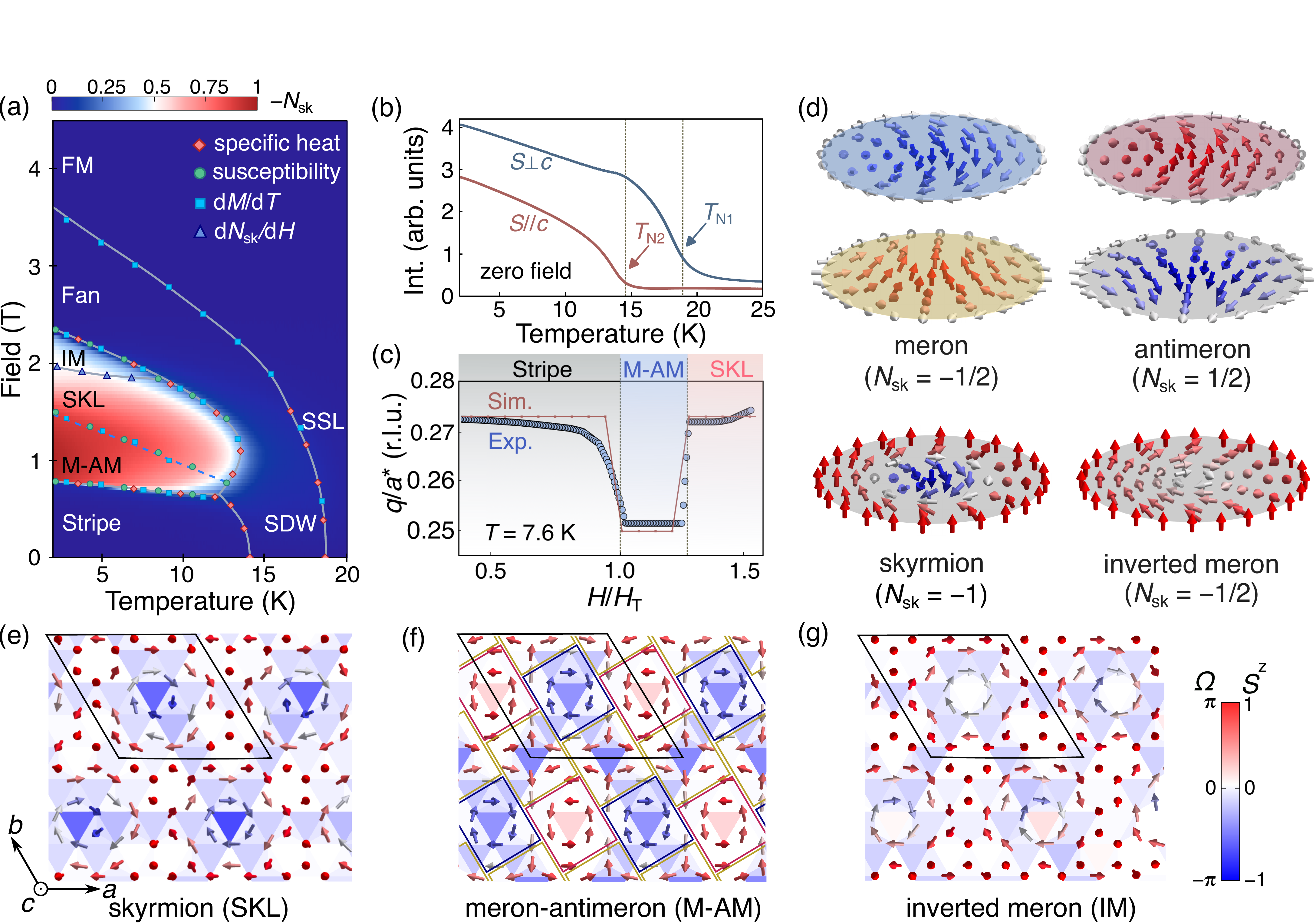}
    \caption{Topological phase transitions in Gd$_3$Ru$_4$Al$_{12}$. (a) Theoretical phase diagram for Gd$_3$Ru$_4$Al$_{12}$ calculated in a magnetic field along the $c$ axis. Calculations were performed for the equivalent triangular lattice model in Table~\ref{tab:exchange} using classical Monte Carlo simulations. Pseudocolor corresponds to the negative skyrmion number, $-N_\mathrm{sk}$. The phase boundaries are determined from the calculated specific heat (red diamonds), magnetic susceptibility (green circles), d$M$/d$T$ (dark blue square) and d$N_\mathrm{sk}$/d$H$ (light blue triangles). The observed phases include the stripe, sinusoidal modulated spin density wave (SDW), spiral spin-liquid (SSL), meron-antimeron (M-AM),  skyrmion, inverted meron (IM), and fan structures. The topological phase transition between the meron-antimeron phase and the skyrmion phase is delineated by a blue dashed line. (b) Temperature evolution of the simulated magnetic structure factor components along the $c$ axis (red) and in the $ab$ plane (blue). (c) Simulated (red line) and experimental (blue circles) field evolution of the magnetic long-range order wavevector $q/a^*$. Calculations were performed on a $44\times 44\times 2$ supercell of the effective triangular lattice. Experimental data are adapted from Ref.~\cite{hirschbergerLatticecommensurate2024}. (d) Prototype topological spin textures, including two types of merons ($N_\mathrm{sk} = -1/2$) and antimerons ($N_\mathrm{sk} = 1/2$), skyrmion ($N_\mathrm{sk} = -1$), and inverted meron ($N_\mathrm{sk} = -1/2$). (e-g) Simulated magnetic structures in the skyrmion phase (e), meron-antimeron phase (f), and inverted meron phase (g) on two successive layers of the equivalent AB-stacked triangular lattice viewed along the $c$ axis. Pseudocolor of the spins corresponds to the length of spin component along the $c$ axis. Pseudocolor in the background corresponds to the solid angle $\mathit{\Omega}$ as defined in the Supplemental Material \cite{supp}. Unit cells of the magnetic structures are outlined by black lines. In panel (f), one magnetic unit cell is composed of three merons and one anti-meron, each components being outlined by rectangles of the same color as those of the prototype spin textures in panel (d). }
    \label{fig:fig4}
\end{figure*} 
\section{Spin dynamics and multi-target modeling}

Inelastic neutron scattering (INS) provides access to magnetic excitations over extended regions of reciprocal space, offering crucial insights into the underlying spin interactions \cite{lovesey_theory_1984}. Figures~\hyperlink{anchor:fig2-top}{\ref{fig:fig2}(a,b)} summarize the INS spectra obtained from co-aligned $^{160}$Gd$_3$Ru$_4$Al$_{12}$ single crystals. These include cuts along high-symmetry directions in momentum space (Fig.~\hyperlink{anchor:fig2-top}{\ref{fig:fig2}(a)}) and constant-energy slices at energy transfers of 1.0 and 8.5~meV (Fig.~\hyperlink{anchor:fig2-top}{\ref{fig:fig2}(b)}). Two distinct magnon branches are observed below the N\'eel temperature ($T = 5$~K), spanning energy ranges of 0--3 and 7--9~meV. These can be attributed to inter-trimer and intra-trimer excitations, respectively. At elevated temperature ($T = 100$~K), only a softened remnant of the high-energy branch remains, indicating the robustness of local trimer correlations due to dominant intra-trimer interactions, even well above $T_\mathrm{N}$.

To understand the magnetic properties of  Gd$_3$Ru$_4$Al$_{12}$, we consider a microscopic model on the original breathing kagome lattice. The spin Hamiltonian can be written as
\begin{equation}
\mathcal{H} = \sum_{ij \in n} J_{n} \mathbf{S}_i \cdot \mathbf{S}_j + \mathcal{H}_\mathrm{dd} +  \sum_j [K_\mathrm{ab} (S^z_j)^2 - g \mu_B H S^z_j]\mathrm{,}
\label{eq:Hmodel}
\end{equation}
where the first term represents the Heisenberg interactions that are mainly mediated by conduction electrons, $\mathcal{H}_\mathrm{dd}$ represents the magnetic dipole-dipole interactions, $K_\mathrm{ab}$ is the strength of the easy-plane SIA, and the last term corresponds to the Zeeman coupling to an external magnetic field $H$. The gyromagnetic factor $g$ can be approximated by $g\simeq 2$ because $J=S=7/2$ for Gd$^{3+}$ ($L=0$). As described in Fig.\hyperlink{anchor:fig1-top}{~\ref{fig:fig1}(a)}, the exchange interactions, $ J_\mathrm{n}$, include the intra-layer couplings $J_1$, $J_2$, and $J_6$, the inter-layer couplings $J_\mathrm{c1}$, $J_\mathrm{c2}$, and $J_\mathrm{c3}$, and the second-layer couplings $J_\mathrm{n1}$ and $J_\mathrm{n2}$. The resulting exchange interactions on the effective triangular and honeycomb lattices are described in Figs.\hyperlink{anchor:fig1-top}{~\ref{fig:fig1}(b)} and \hyperlink{anchor:fig1-top}{(c)}, respectively. Introducing additional exchange interactions, either among the further-neighboring trimers or among the trimers that are already coupled in the $J_{126}$-$J_{\rm{c123}}$-$J_{\rm{n12}}$ model, does not impact the goodness-of-fit. Due to the complexity of the microscopic model, the inverse scattering problem is underconstrained by the INS data and a multi-target fitting approach must be used to extract the coupling strengths $J_{n}$ and $K_\mathrm{ab}$. The fit simultaneously captures diffuse neutron scattering pattern, INS spectra, magnetization data, and FMR data. Further details on the fitting procedure are provided in the Methods section and the Supplemental Material \cite{supp}.

Table~\ref{tab:exchange} summarizes the fitted exchange parameters of the microscopic model on the original kagome lattice, alongside the corresponding effective couplings mapped onto the triangular and honeycomb lattices. The constraints $J_\mathrm{c1} = J_\mathrm{c2}$ and $J_\mathrm{n1} = J_\mathrm{n2}$ are essential for preserving a narrow bandwidth of the intra-trimer excitation. As shown in Figs.~\ref{fig:fig1} and~\ref{fig:fig2}, the calculated diffuse scattering patterns and INS spectra closely match the experimental observations. Additional results, including fits to the magnetization and FMR data, are provided in the Supplemental Material (Figs.~S3--S4) \cite{supp}. The intra-trimer coupling $J_1$ is found to be an order of magnitude stronger than the other exchange interactions, supporting the assumption of rigid ferromagnetic trimers \cite{nakamuraSpin2018, matsumuraHelical2019, hirschbergerSkyrmion2019}. On the effective honeycomb lattice, the frustration ratio between intra- and inter-sublattice couplings is estimated to be $\sim 0.21$, exceeding the critical threshold of $1/6$ required to stabilize spiral spin-liquids \cite{gaoSpiral2022}. These results demonstrate that our microscopic model captures the essential ingredients responsible for the emergence of a codimension-two spiral spin-liquid in Gd$_3$Ru$_4$Al$_{12}$.

\section{Insights into the topological spin textures}

The most compelling validation of our microscopic model, obtained through multi-target fitting, is the depth of understanding it offers into the complex and diverse magnetic phases of Gd$_3$Ru$_4$Al$_{12}$~\cite{hirschbergerSkyrmion2019}.
Figure~\hyperlink{anchor:fig4-top}{\ref{fig:fig4}(a)} shows the phase diagram under a magnetic field applied along the $c$ axis, obtained from classical Monte Carlo simulations on a $12\times 12\times 4$ supercell of the effective triangular lattice. The pseudocolor scale represents the calculated skyrmion number, $N_\mathrm{sk}$. For each field strength, simulations were performed at multiple temperatures using parallel tempering. Consistent with experimental observations \cite{nakamuraSpin2018, matsumuraHelical2019, hirschbergerSkyrmion2019}, two successive phase transitions are seen at zero field: the first at $T_\mathrm{N1} \sim 18$~K and the second at $T_\mathrm{N2} \sim 15$~K. As illustrated in Fig.~\hyperlink{anchor:fig4-top}{\ref{fig:fig4}(b)}, analysis of the magnetic structure factors along different directions reveals that the intermediate phase between $T_\mathrm{N1}$ and $T_\mathrm{N2}$ corresponds to a spin-density wave (SDW) with spins oriented perpendicular to both $\bm{c}$ and $\mathbf q$. Below $T_\mathrm{N2}$, a finite spin component along $\bm{c}$ emerges in agreement with experiments \cite{matsumuraHelical2019, hirschbergerSkyrmion2019}. Aside from the main component of an elliptical spiral order~\cite{matsumuraHelical2019, hirschbergerSkyrmion2019}, our model suggests the coexistence of a secondary component of a sinusoidally modulated collinear order, which leads to a double-${\mathbf q }$ stripe structure like that observed in GdRu$_2$Si$_2$~\cite{khanh_zoology_2022}. As presented in the Supplemental Material (Fig.~S10) \cite{supp}, this stripe order is consistent with our neutron diffraction dataset and the polarization analysis in the resonant x-ray diffraction experiments~\cite{matsumuraHelical2019, hirschbergerSkyrmion2019}.

\begin{figure*}[t]
    \centering
    \hypertarget{anchor:fig6-top}{}
    \includegraphics[width=0.9\linewidth]{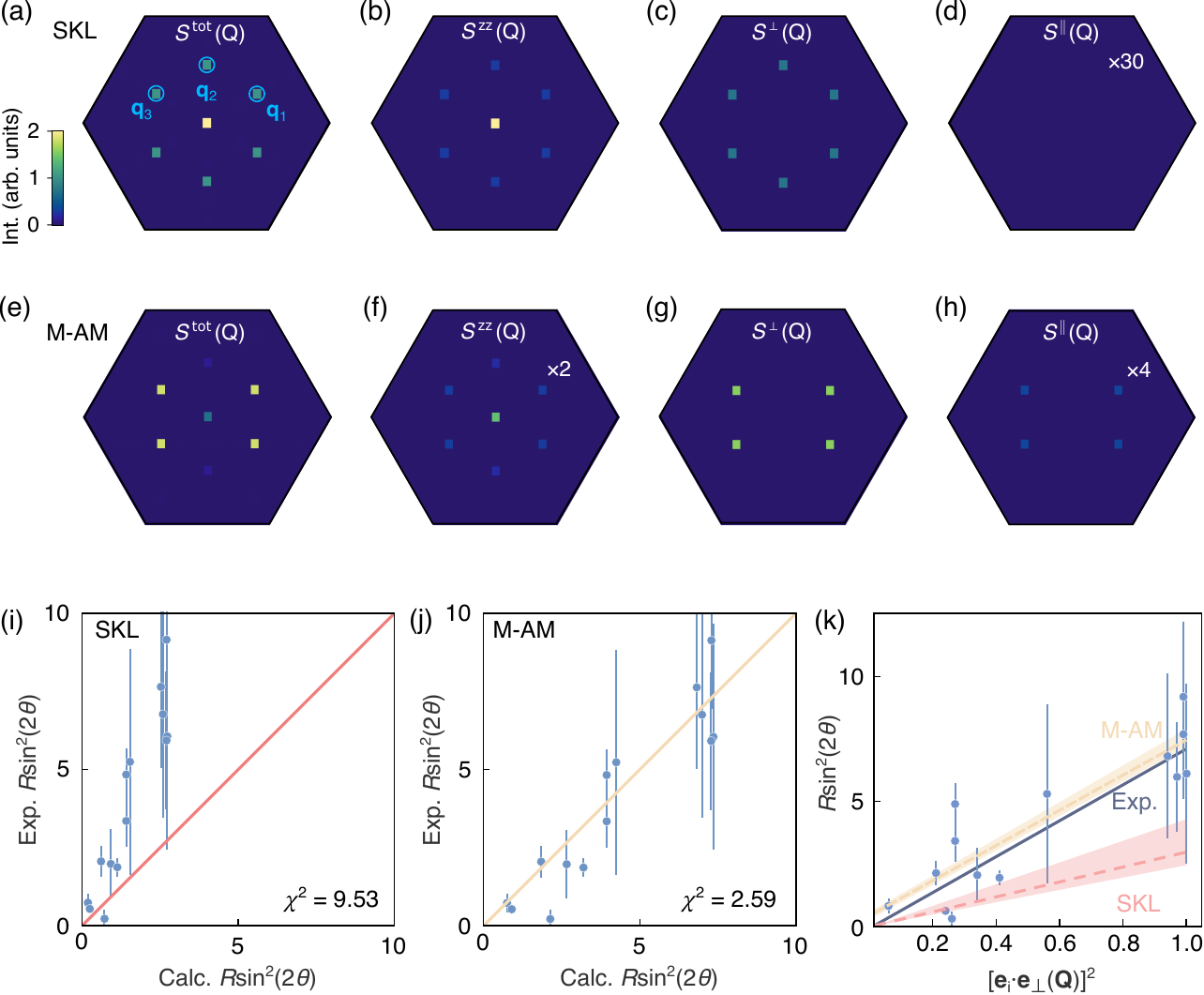}
    \caption{Identification of the commensurate M-AM lattice through comparison with the resonant x-ray diffraction data. (a)-(h) Calculated magnetic structure factor, $S(\mathbf{Q})$, and its decomposition for the SKL in (a)-(d) and the M-AM lattice in (e)-(h). Calculations were performed on the original breathing kagome lattice at zero temperature in a field of 1.0 and 1.7~T for the M-AM and SKL states, respectively. The components shown are the total structure factor $S^\mathrm{tot}(\mathbf{Q})$, the out-of-plane component $S^\mathrm{zz}(\mathbf{Q})$, the in-plane component perpendicular to the wave vector $S^\mathrm{\perp}(\mathbf{Q})$, and the in-plane component parallel to the wave vector $S^\mathrm{\parallel}(\mathbf{Q})$. (i, j) Comparison between the  experimental (blue markers) and calculated values of $R\sin^2(2\theta)$ for the (i) SKL and (j) M-AM states. The experimental data, which were measured in the commensurate phase at $T = 7.8$~K in $H = 1.3$~T, are adapted from Ref.~\cite{hirschbergerLatticecommensurate2024}. Calculations were performed on the original breathing kagome lattice in a field of $H$ = 1.3 T. (k) Ellipsoid plot for the same polarized x-ray diffraction data (blue markers), for which the linear fit is shown by the solid blue line. Dashed yellow and red lines are calculated for the M-AM and SKL models at $H$ = 1.3 T, respectively. The shaded areas describe the variance of the calculated slope within the stable field range of 0.8 to 1.5~T and 1.2 to 2.0~T for the M-AM and SKL states, respectively, with the lower boundary corresponding to the lower field.}
    \label{fig:fig6}
\end{figure*} 

The most prominent feature of the phase diagram in Fig.~\hyperlink{anchor:fig4-top}{\ref{fig:fig4}(a)} is the emergence of a field-induced topological phase persisting down to $\sim2$~K. This agrees with experimental observations under field cooling, though it differs slightly from the higher stabilization temperature of $\sim4$~K observed under zero-field cooling \cite{hirschbergerSkyrmion2019}. As explained in the Supplemental Material~\cite{supp}, this discrepancy may arise from additional terms not included in our model, such as redundant spin interactions among the trimers or longer-range exchange interactions.

To clarify the roles of SIA and dipolar interactions, we selectively remove these terms in the simulations presented in the Supplemental Material (Fig. S12) \cite{supp}. In their absence, magnetic frustration alone produces a narrow skyrmion crystal phase adjacent to the paramagnetic regime—a mechanism reminiscent of that proposed by Okubo, Chung and Kawamura \cite{Okubo2012}. Reintroducing the dipolar interactions significantly broadens the stability range of the topological phase, effectively counteracting the suppressive influence of the easy-plane SIA and yielding a robust topological region as shown in Fig.~\hyperlink{anchor:fig4-top}{\ref{fig:fig4}(a)}.

Despite its relatively weak magnitude, the presence of easy-plane SIA significantly enriches the diversity of spin textures in Gd$_3$Ru$_4$Al$_{12}$. Previous experiments have identified an enigmatic field-induced transition within the topological phase, evident in magnetic susceptibility measurements \cite{hirschbergerSkyrmion2019}. This transition is accompanied by a subtle shift in the propagation vector $\mathbf {q}$, from near (0.25, 0, 0) 
to (0.272, 0, 0) 
 \cite{hirschbergerLatticecommensurate2024}. As shown in Fig.\hyperlink{anchor:fig4-top}{~\ref{fig:fig4}(a)}, this nuanced transition is also captured in our simulations. By using an extended supercell of $44\times 44\times 2$, which is commensurate with both the (1/4, 0, 0) and (3/11, 0, 0) wave vectors, the results presented in Fig.\hyperlink{anchor:fig4-top}{~\ref{fig:fig4}(c)} successfully reproduce the observed shift in $\mathbf {q}$.

Using the prototype topological spin textures in Fig.\hyperlink{anchor:fig4-top}{~\ref{fig:fig4}(d)} as a reference, we analyze the simulated magnetic orders in real space. Above the transition field, the spin configuration shown in Fig.\hyperlink{anchor:fig4-top}{~\ref{fig:fig4}(e)} corresponds to a triangular lattice of magnetic skyrmions with topological charge $N_\mathrm{sk} = -1$. Below the transition, the texture depicted in Fig.\hyperlink{anchor:fig4-top}{~\ref{fig:fig4}(f)} displays fractional topological charges of $N_\mathrm{sk} = -1/2$ and $1/2$, forming a novel meron–antimeron lattice reminiscent of that observed in the non-centrosymmetric compound Co$_8$Zn$_9$Mn$_3$~\cite{yuTransformation2018}. Each magnetic unit cell contains three merons and one antimeron, resulting in a total topological charge of $N_\mathrm{sk} = -1$, matching that of the skyrmion lattice.

The difference in the topological spin textures across the commensurate phase transition can also be resolved in the calculated magnetic structure factors, which are presented in Figs.~\ref{fig:fig6}(a)-\ref{fig:fig6}(d) and \ref{fig:fig6}(e)-\ref{fig:fig6}(h) for the skyrmion and meron-antimeron states, respectively. For the skyrmion lattice, the $\mathcal{S}^\mathrm{zz}(\mathbf{Q})$ and $\mathcal{S}^{\bot}(\mathbf{Q})$ components, which are the out-of-plane component and the in-plane component perpendicular to $\mathbf{Q}$, respectively, exhibit a characteristic triple-${\mathbf q}$ pattern \cite{muhlbauerSkyrmion2009}, while the in-plane component parallel with $\mathbf{Q}$, $S^\parallel(\mathbf{Q})$, exhibits zero intensity, as is consistent with a helical component for all three arms. 
In contrast, in the meron-antimeron phase, 
the $\mathcal{S}^\mathrm{zz}(\mathbf{Q})$ and $\mathcal{S}^{\bot}(\mathbf{Q})$  components exhibit triple-${\mathbf q}$ and double-${\mathbf q}$ patterns, respectively. The $S^\parallel(\mathbf{Q})$ component, which becomes non-zero in the meron-antimeron state, follows the same double-${\mathbf q}$ pattern as that of the $\mathcal{S}^{\bot}(\mathbf{Q})$  component.

Given its unique magnetic structure factor, the meron-antimeron lattice can be directly verified through polarized resonant x-ray diffraction. For experiments with ($h$, $k$, 0) as the horizontal scattering plane, magnetic structure factors along the $\mathbf{Q}$ vectors and the $\bm{c}$ axis can be distinctively probed in the $\pi\to\sigma'$ and $\pi\to\pi'$ channels, respectively. Here, the incident x-ray polarization ($\pi$) is parallel to the scattering plane, while the scattered x-ray polarization is either ($\pi'$) parallel with  or ($\sigma'$)  perpendicular to the plane. More detailed description of this experiment can be found in the Supplemental Material~\cite{supp}.

For the commensurate magnetic Bragg peaks in the topological phase of Gd$_3$Ru$_4$Al$_{12}$, the intensity ratio between the $\pi\to\sigma'$ and $\pi\to\pi'$ channels, $R=I_{\pi\to \sigma'}/I_{\pi\to\pi'}$, has already been tabulated in the Supplementary Table I of Ref.~\cite{hirschbergerLatticecommensurate2024}. Denoting  the scattering angle as $2\theta$, Figs~\ref{fig:fig6}(i) and (j) compare the experimental values of $R\sin^2(2\theta)$ against the calculated values for the simulated meron-antimeron and skyrmion structures~\cite{supp}. This comparison reveals that the meron-antimeron phase is in excellent agreement with the experimental data, yielding a goodness-of-fit factor $\chi^2=2.59$, whereas a significant deviation of $\chi^2=9.53$ is observed for the skyrmion structure.

Insight for the stabilization of the meron-antimeron lattice can be further drawn from the ellipticity plot presented in Fig.~\ref{fig:fig6}(k), in which $R\sin^2(2\theta)$ is plot against $\mathbf [\mathbf e_\mathrm i \cdot \mathbf e_\perp (\mathbf Q)]^2$. Here $\mathbf e_\mathrm i$ denotes the unit vector along the incident x-ray direction and $\mathbf e_\perp (\mathbf Q)$ denotes the in-plane unit vector perpendicular to $\mathbf Q$. In such a plot, the slope is proportional to the ellipticity,  $\mathcal{S}^{\bot}(\mathbf{Q})/\mathcal{S}^\mathrm{zz}(\mathbf{Q})$ ~\cite{supp}. Therefore, the relatively high slope for the meron-antimeron structure, which closely matches the experimental value and is much higher than the values for the skyrmion, reveals an enhanced ellipticity in the meron-antimeron lattice. Similar conclusion can be drawn from the relatively weak $\mathcal{S}^\mathrm{zz}(\mathbf{Q})$ component of the meron-antimeron lattice shown in Fig.~\ref{fig:fig6}(f). This enhanced ellipiticity confirms that the meron-antimeron lattice is stabilized by the easy-plane SIA, similar to the scenario in the noncentrosymmetric systems~\cite{Lin2015}.

In addition to the meron-antimeron lattice, our simulations also suggest the emergence of a second type of meron lattice in the transitional regime between the skyrmion and fan phases. This spin texture, illustrated in Fig.\hyperlink{anchor:fig4-top}{~\ref{fig:fig4}(g)}, is completely inverted relative to that of conventional merons \cite{su_crystallized_2012}. Given its narrow stability range, further experimental validation is needed to confirm the existence of this inverted meron phase.

\begin{figure*}[t]
    \centering
    \hypertarget{anchor:fig5-top}{}
    \includegraphics[width=1\linewidth]{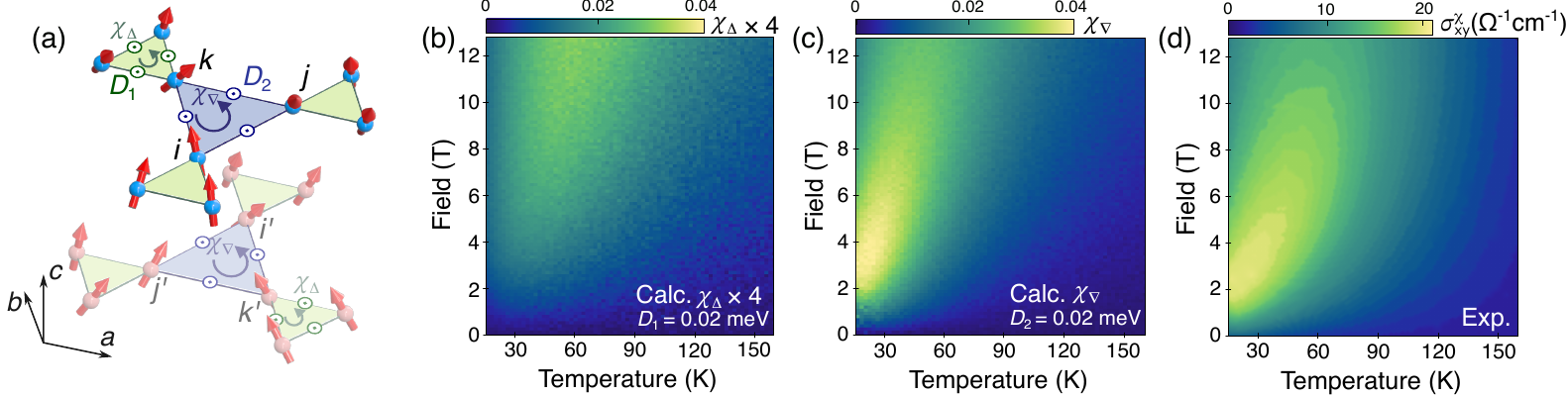}
\caption{Chiral spin fluctuations in the spiral spin-liquid regime. (a) Scalar spin chirality, $\chi_\Delta$ and $\chi_\nabla$, over the small and large triangles on the original breathing kagome lattice, respectively. DM interactions with strengths of $D_1$ and $D_2$ along the $c$ axis are indicated by blue and green circled  dots, respectively. (b) Temperature and field-dependence of $\chi_\Delta$ calculated for the fitted $J_{126}$-$J_{\rm{c123}}$-$J_{\rm{n12}}$ model with $D_1 = 0.02$~meV and $D_2 = 0$. (c) Similar calculations for $\chi_\nabla$ with $D_1=0$ and $D_2 = 0.02$~meV. (d) Experimental chirality-driven Hall conductivity, $\sigma_{xy}^{\chi}$, as reproduced from the previous report~\cite{kolincioKagome2023}.}
    \label{fig:fig5}
\end{figure*} 
\section{Chiral fluctuations in spiral spin-liquids}

Unlike previously known experimental realizations, the spiral spin-liquid in Gd$_3$Ru$_4$Al$_{12}$ exhibits strong chiral fluctuations \cite{kolincioKagome2023}, suggesting additional perturbations that, when combined with the  Zeeman term, can induce a net chirality. Building on prior discussions of spin trimers \cite{kolincio_large_2021, kolincioKagome2023}, we now examine the effects of DM interactions, while ensuring that their weak strengths do not significantly impact the fit to the experimental data (see Fig.~S17 in the Supplemental Material \cite{supp}).

According to Fig.~\hyperlink{anchor:fig5-top}{\ref{fig:fig1}(e)}, the staggered DM interactions, dictated by the centrosymmetry of Gd$_3$Ru$_4$Al$_{12}$, suggest opposite vector chiralities in neighboring layers. As described in Fig.~\hyperlink{anchor:fig5-top}{\ref{fig:fig5}(a)}, for the scalar spin chirality $\chi_{ijk} = (\mathbf{S}_{i} \times \mathbf{S}_{j}) \cdot \mathbf{S}_{k}$ on the triangle $ijk$ in counterclockwise circulation, the inversion center located between the neighboring layers also flips the ordering of sites along each bond of the associated triangular plaquettes. Consequently, a net scalar spin chirality—detectable via the anomalous Hall effect \cite{nagaosa_anomalous_2010}—can still emerge from the DM interactions, despite the global inversion symmetry.

Figures\hyperlink{anchor:fig5-top}{~\ref{fig:fig5}(b,c)} illustrate the separate impacts of perturbative DM interactions, including $D_1$ ($D_2$) over the $J_1$ ($J_2$) bonds, on the chirality over the smaller and larger triangles, defined as  $\chi_\Delta$ and  $\chi_\nabla$, respectively. The experimental anomalous Hall conductivity, $\sigma_{xy}^{\chi}$, is reproduced in Fig.\hyperlink{anchor:fig5-top}{~\ref{fig:fig5}(d)} for comparison. Although both $\chi_\Delta$ and $\chi_\nabla$ become nonzero in the presence of the DM interactions, their temperature and field evolutions are very different. In the investigated regime, $\chi_\Delta$ is steadily enhanced by magnetic field at elevated temperatures, while a maximum in $\chi_\nabla$ is observed in the field regime of 2-4~T at lower temperatures. As discussed in the Supplemental Material \cite{supp}, these contrasting behaviors are robust against variations in the DM interaction strengths and thus reflect the intrinsic disparity between $J_1$ and $J_2$ revealed by our microscopic model. Given the strong resemblance between the field dependence of $\sigma_{xy}^{\chi}$ and that of $\chi_\nabla$, we conclude that the anomalous Hall effect observed in the spiral spin-liquid phase of Gd$_3$Ru$_4$Al$_{12}$ predominantly originates from chiral fluctuations on the larger triangles.

\section{Discussion}
     
The diverse chiral orders revealed in our study of Gd$_3$Ru$_4$Al$_{12}$ carry important implications for both the fundamental understanding and potential applications of topological spin textures. A key outcome is the establishment of a successful framework for constructing realistic and predictive microscopic models for intermetallic frustrated magnets. This represents a crucial breakthrough, as previous attempts to model this class of materials microscopically often struggled to reproduce key experimental observations across the full phase diagram, while phenomenological models~\cite{kezsmarki2015neel, chacon2018observation, takagi2022square} offer limited physical insight and predictive power compared to a microscopic model constrained by sufficiently comprehensive datasets. By utilizing the state-of-the-art optimization algorithms or machine learning techniques~\cite{Samarakoon2020}, the multi-target approach can provide a systematic and scalable method to extract the model and physics underlying the topological spin textures and frustrated magnets in general.

As demonstrated in our work, the predictive power of our microscopic model allows us to provide quantitative and definitive confirmation that magnetic frustration is the dominant stabilization mechanism for the skyrmion lattice in Gd$_3$Ru$_4$Al$_{12}$. This approach unambiguously eliminates other potential mechanisms such as multiple spin interactions~\cite{hirschbergerNanometric2021, hayami_stabilization_2024} or $d$-$f$ inter-orbital competitions~\cite{nomoto_formation_2020, Takuya2023}. While frustration-based scenarios have been proposed for other Gd-based compounds like Gd$_2$PdSi$_3$~\cite{bouazizFermiSurface2022, paddisonMagnetic2022, dongFermi2024} and GdRu$_2$Si$_2$~\cite{bouazizFermiSurface2022, paddisonSpin2024, woodMagnon2025a}, our work provides the first rigorous validation through a realistic microscopic model, and suggests this mechanism may be broadly applicable to other centrosymmetric skyrmion hosts.

Further insights from our study emerge from the distinctive role played by the easy-plane SIA, which has traditionally been regarded as detrimental to the stabilization of magnetic skyrmions~\cite{wangMeron2021a,leonovMultiply2015}. Contrary to this prevailing view, our results demonstrate that the easy-plane SIA in Gd$_3$Ru$_4$Al$_{12}$ significantly enriches the landscape of topological spin textures by stabilizing an intermediate inverted meron lattice and a robust meron-antimeron lattice~\cite{Lin2015}. Notably, previous experimental realizations of the meron-antimeron lattice~\cite{yuTransformation2018} have been limited to non-centrosymmetric materials, where the large pitch of the spin texture suppresses the emergence of a substantial anomalous Hall effect. In contrast, the atomic-scale meron-antimeron lattice in Gd$_3$Ru$_4$Al$_{12}$ establishes it as a unique platform for exploring magneto-electric responses associated with fractional topological charges. Furthermore, advances in interface engineering, which enable precise tuning of the type and strength of SIA~\cite{dieny_perp_2017}, could allow controlled manipulation of topological charge fractionalization and coalescence in this compound, expanding the functionality of skyrmion-based spintronic devices.

\section{Methods}
\subsection{Sample preparation}
Polycrystals of $^{160}$Gd$_3$Ru$_4$Al$_{12}$ were synthesized by arc melting in argon atmosphere, where the $^{160}$Gd-enriched isotope was used to reduce neutron absorption. The obtained polycrystals were shaped into rods using the arc melting method. The single crystal of  $^{160}$Gd$_3$Ru$_4$Al$_{12}$ was grown under argon gas flow by a floating zone furnace from the rods. At an incident neutron energy of $E_\mathrm i = 22$~meV, the neutron absorption cross section of the $^{160}$Gd$_3$Ru$_4$Al$_{12}$ crystal is estimated to be $\sim7383$~bn/f.u., leading to a penetration depth of $\sim0.9$~mm. The samples were characterized by powder x-ray diffraction (XRD) and energy-dispersive x-ray spectroscopy (EDX).

\subsection{Neutron diffraction experiments}
Neutron diffraction experiments were performed on the WAND$^2$ diffractometer \cite{frontzek_wand_2018} at the High Flux Isotope Reactor (HFIR), Oak Ridge National Laboratory (ORNL). An incident neutron wavelength of $\lambda = 1.486$~\AA\ was selected by a Ge-113 monochromator. A closed cycle refrigerator (CCR) was utilized to access temperatures down to 4.2~K. A piece of $^{160}$Gd$_3$Ru$_4$Al$_{12}$ crystal with a mass of $\sim 60$~mg was aligned with the ($h$,~0,~$l$) plane horizontal. The crystal is of a plate shape, with a thickness of $\sim1$~mm along the $\bm{c}$ axis to optimize the scattering intensity. Measurements were performed at $T = 4.2$, 20, and 300~K, by rotating the sample around the vertical axis in 0.1$^\circ$ steps, covering a total range of 180$^\circ$. The counting time at each step is 30, 60, and 30~s for measurements at $T = 4.2$, 20, and 300~K, respectively. From the $T = 4.2$~K measurements, 92 magnetic reflections were subtracted using the MANTID package \cite{arnold_mantid_2014}, and their intensities were corrected for neutron absorption before being compared to the calculations. To obtain the diffuse scattering patterns at $T = 20$~K shown in Fig.\hyperlink{anchor:fig1-top}{~\ref{fig:fig1}}, data at $T = 300$~K were subtracted as the background, with the scattering intensity being corrected for neutron absorption.

\subsection{Inelastic neutron scattering experiments}
INS experiments were performed on the SEQUOIA spectrometer \cite{ehlers_new_2011} at the Spallation Neutron Source (SNS), ORNL. An incident neutron energy of $E_\mathrm{i} = 22$~meV was used in the high resolution mode with a Fermi chopper frequency of 240 Hz. A CCR was utilized to access temperatures down to 5~K. 8 pieces of $^{160}$Gd$_3$Ru$_4$Al$_{12}$ crystals, with a total mass of $\sim0.52$~g, are coaligned on an aluminum plate with the ($h$,~0,~$l$) plane horizontal. The crystals are all of a plate shape with a thickness of $\sim1$~mm along the $\bm{c}$ axis. Measurements at $T = 5$~K (100~K) were performed by rotating the sample around the vertical axis in 0.5$^\circ$ (1$^\circ$) steps, covering a total range of 134$^\circ$. Data reduction and projection were performed using the MANTID \cite{arnold_mantid_2014} and HORACE \cite{ewings_horace_2016} packages.  
 
\subsection{Multi-target analysis}
Multi-target fits were performed to achieve a comprehensive understanding of the experimental data, which include the diffuse neutron scattering patterns, INS spectra, magnetization at characteristic measuring fields, and angle dependence of the FMR fields. Details of the input data are presented in the Supplemental Material \cite{supp}. Using the spin Hamiltonian as shown in Eq.~\ref{eq:Hmodel}, the diffuse neutron scattering pattern and the magnetization were calculated by the classical Monte Carlo simulations with Langevin updates, the INS spectra and the FMR fields were calculated by the Landau-Langevin method, both algorithms being implemented in the Sunny.jl program~\cite{dahlbomSunnyjl2025s}. The cost function was optimized using the Marine Predators Algorithm (MPA)~\cite{FARAMARZI2020113377} with 96 search agents, each iterating 100 times to reach the minimal $\chi^2$ values. 
  
\subsection{Classical Monte Carlo simulations for long- and short-range chiral orders}
Classical Monte Carlo simulations with Langevin dynamics, combined with the parallel tempering technique as implemented in the Sunny.jl package (v0.7)~\cite{dahlbomSunnyjl2025s}, were employed to obtain the phase diagram, determine the spin textures, and calculate the DM-induced chiral fluctuations in Gd$_3$Ru$_4$Al$_{12}$. Unless otherwise stated, simulations were performed on a $12\times 12 \times 4$ supercell of the original breathing kagome lattice model with periodic boundary conditions. The phase diagram was obtained by calculating the skyrmion number, magnetic susceptibility, and specific heat. More details for the calculations can be found in the Supplemental Material \cite{supp}.

\begin{acknowledgments} 
We wish to acknowledge helpful discussions with Max Hirschberger, Licong Peng, Zhentao Wang, Yang Gao, James Jun He, Hao Zhang, David A. Dahlbom, Gang Chen, Yuan Li, and Otkur Omar. Work at the University of Science and Technology of China was supported by National Key R\&D Program of China under the Grant
No.~2024YFA1613100 and the National Natural Science Foundation of China (NSFC) under the Grant No.
12374152. Work at the University of Tennessee was supported by National Science Foundation
Materials Research Science and Engineering Center program through the UT Knoxville Center for
Advanced Materials and Manufacturing (DMR-2309083). Works at RIKEN were supported by
JST CREST (Grant No. JPMJCR20T1), and the RIKEN TRIP initiative (Many-body Electron Systems 
and Advanced General Intelligence for Science Program). Work at ORNL  was  supported  by  the  U.S.  Department  of Energy,  Office  of  Science,  Basic  Energy  Sciences,  Materials  Science  and  Engineering  Division.  A portion of this research used resources at the High Flux Isotope Reactor and Spallation Neutron Source, DOE Office of Science User Facilities operated by the Oak Ridge National Laboratory.  Beam time was allocated to WAND$^2$ on proposal number IPTS-29966, and to \mbox{SEQUOIA} on proposal number IPTS-29962. 
\end{acknowledgments}



%

\setlength{\bibsep}{1em}%

\clearpage
\newpage

\renewcommand{\thefigure}{S\arabic{figure}}
\renewcommand{\thetable}{S\arabic{table}}
\renewcommand{\theequation}{S\arabic{equation}}

\makeatletter
\renewcommand*{\citenumfont}[1]{S#1}
\renewcommand*{\bibnumfmt}[1]{[S#1]}
\def\clearfmfn{\let\@FMN@list\@empty}  
\makeatother
\clearfmfn

\setcounter{figure}{0} 
\setcounter{table}{0}
\setcounter{equation}{0} 
\setcounter{section}{0} 

\onecolumngrid
\begin{center} {\bf \large Supplemental Materials for:\\Skyrmion and Meron Crystals in Intermetallic Gd$_3$Ru$_4$Al$_{12}$: Microscopic Model Insights into Chiral Phases} \end{center}

\vspace{0.5cm}

\renewcommand*{\thefootnote}{\arabic{footnote}}
\renewcommand{\thefigure}{S\arabic{figure}}
\renewcommand{\thetable}{S\arabic{table}}
\renewcommand*{\thefootnote}{\arabic{footnote}}

\section{Magnetic susceptibility of the isotope-enriched $^{160}$G\lowercase{d}$_3$R\lowercase{u}$_4$A\lowercase{l}$_{12}$ crystal}
\begin{figure*}[h]
    \centering
\renewcommand{\thefigure}{S\arabic{figure}}
\includegraphics[width=0.4\linewidth]{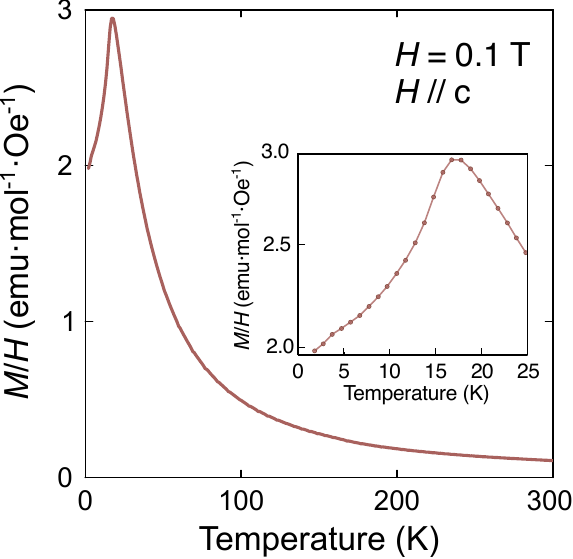}
\hypertarget{anchor:M_raw}{}
    \caption{Magnetic susceptibility of the isotope-enriched $^{160}$Gd$_3$Ru$_4$Al$_{12}$ crystal. Data were measured with a 0.1~T field along the $c$ axis in field cooling. Inset is a zoomed-in plot for comparison with the previous reports~\cite{hirschbergerSkyrmion2019s}}.
    \label{fig:M_raw}
\end{figure*}

Figure~\hyperlink{anchor:M_raw}{\ref{fig:M_raw}} presents the temperature evolution of the magnetic susceptibility measured on a piece of isotope-enriched $^{160}$Gd$_3$Ru$_4$Al$_{12}$ crystal. According to the previous reports on the  Gd$_3$Ru$_4$Al$_{12}$ crystal synthesized with natural abundance Gd~\cite{nakamuraSpin2018s, hirschbergerSkyrmion2019s, nakamuraMagnetic2023}, for the two successive magnetic transitions at $T_\mathrm{N1}=17.2$ and $T_\mathrm{N2}=18.6$~K, a sharp drop in magnetic susceptibility appears at $T_\mathrm{N1}$, but only a weak slope change is discerned at $T_\mathrm{N2}$. Our experimental data presented in Fig.~\ref{fig:M_raw} is consistent with the previous reports~\cite{nakamuraSpin2018s, hirschbergerSkyrmion2019s, nakamuraMagnetic2023}.

\section{Analysis of the neutron scattering data}

Under the dipole approximation, the double differential neutron scattering cross section in magnetic systems can be expressed as:
\begin{equation}
\renewcommand{\theequation}{S\arabic{equation}} 
    \frac{d^2\sigma({\bf q},\omega)}{d\omega d\Omega} 
=\left(\frac{\gamma_\mathrm{n} r_0}{2}\right)^2 \left(\frac{q_f}{q_i}\right)\sum_{\alpha,\beta} \left(\delta_{\alpha,\beta} - \frac{q^{\alpha}q^{\beta}}{q^2}\right) {\cal S}^{\alpha\beta}({\bf q},\omega),
\end{equation} 
where ${\bf q} = {\bf q}_i - {\bf q}_f$ is the momentum transfer to the sample, $\omega$ is the energy transfer, $\Omega$ is the solid angle, and $r_0$ is the classical electron radius, with $\gamma_\mathrm{n} r_0/2 \approx 2.69 \times 10^{-5}$~\AA. 
${\cal S}^{\alpha\beta}({\bf q},\omega)$ is the dynamic spin structure factor (DSSF)
\begin{equation}
\renewcommand{\theequation}{S\arabic{equation}} 
    {\cal S}^{\alpha\beta}({\bf q},\omega) = \frac{1}{2\pi} \int_{-\infty}^{\infty} e^{-i\omega t} \langle S^{\alpha\dag}_{{\bf q}}(0) S^{\beta}_{{\bf q}}(t) \rangle ~dt,
\end{equation}
where  $S^{\alpha}_\mathbf{q}(t)$ with $\alpha = x, y, z$ is the Fourier transform of the real space spin field
\begin{equation}
\renewcommand{\theequation}{S\arabic{equation}} 
    S^{\alpha}_\mathbf{q}(t) = \sum_j e^{-i\mathbf{q}\cdot \mathbf{R}_j}S_{j}^\alpha(t).
\end{equation}
In this expression, $\mathbf{R}_j$ is the position of the $j$-th magnetic site in the lattice, $S_{j}^\alpha(t)$ is the $\alpha$-component of the spin operator at site $j$ at time $t$.

For the analysis of the inelastic neutron scattering (INS) spectra, the time evolution of the spin configuration is computed using the Landau-Lifshitz-Gilbert (LLG) equation \cite{PhysRevLett.53.2497}, as implemented in the \texttt{Sunny.jl} package \cite{dahlbomSunnyjl2025}:
\begin{equation}
\renewcommand{\theequation}{S\arabic{equation}} 
    \frac{d {\bf M}}{dt} = \gamma\, {\bf M} \times \mathbf{H}_\mathrm{eff} - \lambda\, {\bf M} \times \left({\bf M} \times \mathbf{H}_\mathrm{eff}\right),
    \label{Eq:llg}
\end{equation}
where $\gamma = g\mu_\mathrm{B}/\hbar$ is the gyromagnetic ratio, and $\lambda$ is a phenomenological damping constant.  
The effective magnetic field $\mathbf{H}_\mathrm{eff}$ includes all relevant interactions and is given by
\begin{equation}
\renewcommand{\theequation}{S\arabic{equation}} 
    \mathbf{H}_\mathrm{eff} = \mathbf{H}_\mathrm{ext} + \mathbf{H}_\mathrm{ani} + \mathbf{H}_\mathrm{ex},
    \label{Eq:H_eff}
\end{equation}
where $\mathbf{H}_\mathrm{ext}$ is the applied external field, $\mathbf{H}_\mathrm{ani}$ arises from single-ion anisotropy, and $\mathbf{H}_\mathrm{ex}$ accounts for exchange interactions, including magnetic dipole-dipole interactions (DDI).  

The initial spin configurations used in the dynamical simulations were generated via classical Monte Carlo simulations. In this approach, trial updates of individual spins are proposed according to a stochastic version of the LLG equation:
\begin{equation}
\renewcommand{\theequation}{S\arabic{equation}} 
    \frac{d\mathbf{M}}{dt} = \xi\, \mathbf{M} \times \mathbf{H}_\mathrm{eff} - \lambda\, {\bf M} \times \left({\bf M} \times \mathbf{H}_\mathrm{eff} \right),
\end{equation}
where $\xi$ is Gaussian white noise with magnitude $\sqrt{2k_\mathrm{B}T\lambda}$, determined by the fluctuation-dissipation theorem. Here, $k_\mathrm{B}$ is the Boltzmann constant and $T$ the temperature.

For the spin dynamics calculations at $T = 5$ and $100$~K, a system of $N_\mathrm{cells} = 11 \times 11 \times 10$ unit cells was thermalized using Langevin dynamics with a time step of $\Delta t = 0.014~\text{meV}^{-1}$. The system was initialized in the minimum energy configuration obtained via the conjugate gradient method and subjected to $500{,}000$ Langevin updates, with spins updated every 100 iterations. The coupling to the thermal bath was set to $\lambda = 0.2$, chosen empirically following Ref.~\cite{kimBonddependent2023}. A thermalization time of $14~\text{meV}^{-1}$ was found sufficient to reach equilibrium. 

Following thermalization, the spin dynamics were evolved for an additional duration of $280~\text{meV}^{-1}$ using a time step of $2\Delta t$. From this trajectory, $200$ decorrelated spin configurations were extracted and averaged to compute the DSSF. The resulting INS spectra were convoluted with the instrumental energy resolution, modeled as an energy-dependent Gaussian function with full width at half maximum (FWHM) given by
\begin{equation}
\renewcommand{\theequation}{S\arabic{equation}} 
    \mathrm{FWHM}(\omega) = 2.0013 \times 10^{-6} \, \omega^3 + 2.6457 \times 10^{-4} \, \omega^2 - 0.016789 \, \omega + 0.54653~\text{meV},
\end{equation}
for $E_\mathrm{i} = 22$~meV.

Diffuse neutron scattering experiments performed in the correlated paramagnetic regime probe the equal-time pair correlation function
\begin{equation}
\renewcommand{\theequation}{S\arabic{equation}} 
{\cal S}^{\alpha\beta}(\mathbf{q}) = \int_{-\infty}^{\infty} {\cal S}^{\alpha\beta}(\mathbf{q}, \omega) d(\hbar\omega).
\label{Eq:Sab}
\end{equation}

To analyze the diffuse neutron scattering data, classical Monte Carlo simulations with Langevin dynamics were performed on systems of size $N_\mathrm{cells} = 22 \times 22 \times 2$ and $22 \times 2 \times 20$ for calculations in the ($h$, $k$, 1) and ($h$, 0, $l$) scattering planes, respectively. Each simulation consisted of $10^4$ thermalization sweeps followed by $2 \times 10^4$ measurement sweeps. From these, 400 decorrelated spin configurations were extracted and averaged to compute the equal-time pair correlation function. 
To incorporate the effects of instrumental resolution, the simulated data were convoluted with a two-dimensional Gaussian function. The FWHM values were set to 0.038 and 0.149~r.l.u. along the $(h, 0, 0)$ and $(-k, 2k, 0)$ directions, respectively. These values were determined by fitting the profiles of nuclear Bragg peaks at the center of the Brillouin zone.

\section{Analysis of the ferromagnetic resonance (FMR) spectra} 

\begin{figure*}[t] 
    \centering
\renewcommand{\thefigure}{S\arabic{figure}}
    \includegraphics[width=0.8\linewidth]{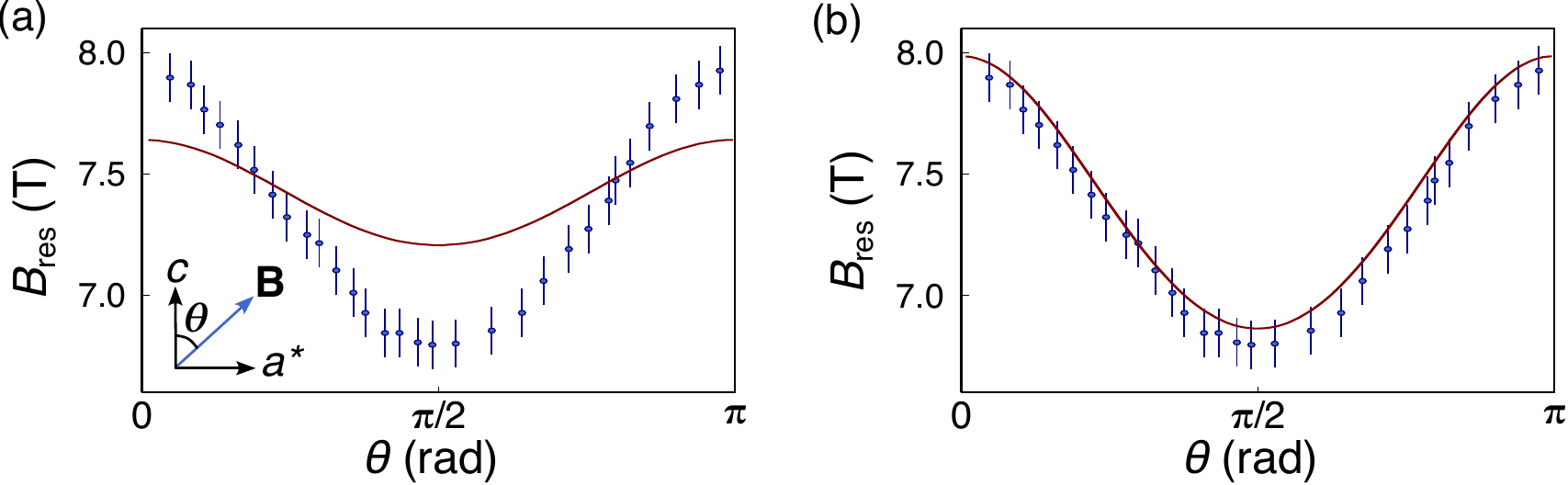} 
    \hypertarget{anchor:ESR}{}
    \caption{Angle dependence of the ferromagnetic resonance field. (a,b) Circular data points are the previously reported experimental resonance field~\cite{hirschbergerLatticecommensurate2024s} measured at $T=2$~K. The rotation angle, $\theta$, is defined in the $a^*c$ plane as shown in the inset of (a). Solid line corresponds to the simulated resonance field using the fitted $J_{126}$-$J_{\rm{c123}}$-$J_{\rm{n12}}$ model without (a) and with (b) an easy-plane single-ion anisotropy of $K_\mathrm{ab} = 0.009$~meV.} 
    \label{fig:ESR} 
\end{figure*}

FMR experiments on Gd$_3$Ru$_4$Al$_{12}$ were conducted in Ref.~\cite{hirschbergerLatticecommensurate2024s} to investigate the single-ion anisotropy (SIA) of the Gd$^{3+}$ spins. In these measurements, the resonance frequency $\omega_\mathrm{res}$ was fixed, and is related to the magnon energy at zero wavevector via
\begin{equation}
\renewcommand{\theequation}{S\arabic{equation}} 
    \hbar \omega_\mathrm{res} (\theta) = E_{\mathbf{k}=0}(\theta),
\end{equation}
where $E_{\mathbf{k}}(\theta)$ denotes the single-magnon dispersion above the saturation field, and $\theta$ is the angle between the applied magnetic field and the $c$ axis, defined in the $a^*c$ plane as shown in the inset of Fig.~\hyperlink{anchor:ESR}{\ref{fig:ESR}}.
As reproduced in Fig.~\hyperlink{anchor:ESR}{\ref{fig:ESR}}, the FMR data support the presence of easy-plane single-ion anisotropy, described by the Hamiltonian $\mathcal{H}_\mathrm{ani} = K_\mathrm{ab}\cos^2(\theta)$, with the anisotropy constant estimated as $K_\mathrm{ab} = 0.13/S^2 = 0.0106$~meV in Ref.~\cite{hirschbergerLatticecommensurate2024s}. Here, we propose that a substantial part of the observed anisotropy originates from DDI \cite{nakamuraMagnetic2023, nakamuraMagnetic2024}.

To support this interpretation, we compute the resonance energy as a function of field orientation using the LLG method for models incorporating both DDI and SIA. Notably, for a magnetic field applied along the $c$ axis, the system retains U(1) symmetry, allowing us to derive an analytical expression for $E_{\mathbf{k}=0}(\theta = 0)$ by solving a tight-binding problem:
\begin{equation}
\renewcommand{\theequation}{S\arabic{equation}} 
    E_{\mathbf{k}= 0} (\theta = 0)
= g\mu_\mathrm{B} (\mu_0 H_\mathrm{ext} - \mu_0 NM_\mathrm{sat}) - S\sum_j J_{ij} - E_\mathbf{D}(\theta = 0) - K_\mathrm{ab}(2S-1).
\end{equation}
In this expression, the first term corresponds to the resonance field, $B_\mathrm{res} = \mu_0H_\mathrm{ext} - \mu_0NM_\mathrm{sat}$, where $\mu_0 N M_\mathrm{sat} = 0.13$~T accounts for the demagnetization field. The saturation magnetization density is $M_\mathrm{sat} = 7~\mu_\mathrm{B}/\text{Gd}^{3+} \approx 0.0037$~meV/T\AA$^3$, and the demagnetization factor $N = 0.33$ is estimated based on the sample geometry. The $g$-factor is taken to be $g = 2.005$, following Ref.~\cite{hirschbergerLatticecommensurate2024s}. The term $E_\mathbf{D}(\theta = 0)$ denotes the dipolar energy for the field-aligned ferromagnetic spin configuration.

\begin{figure*}[t!]
    \centering
\renewcommand{\thefigure}{S\arabic{figure}}
    \includegraphics[width=0.6\linewidth]{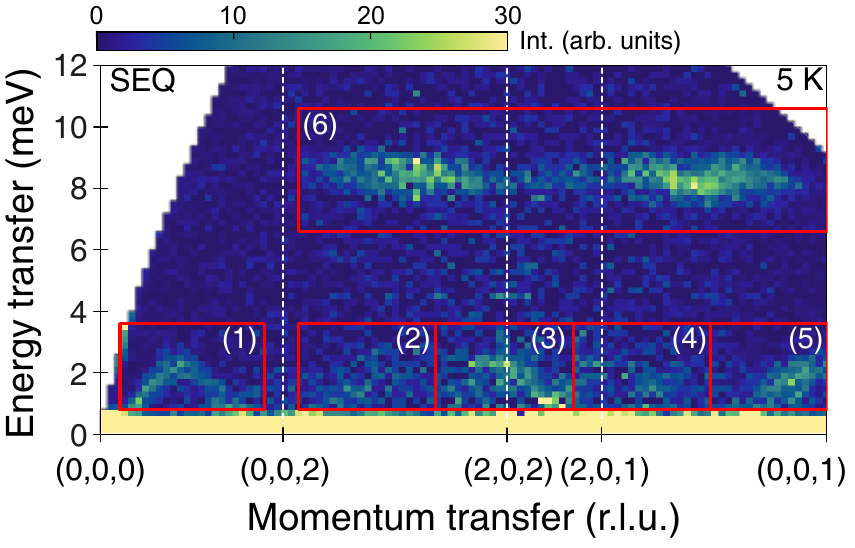}
    \hypertarget{anchor:INS_fit}{}
    \caption{Division of the INS spectra for fits. In the fitting processes, the INS spectra were segmented into six regions as outlined by red boxes. An intensity coefficient was introduced within each segment to compensate for neutron absorption.}
    \label{fig:INS_fit}
\end{figure*}

\section{Multi-target fits}
Multi-target fits were performed to achieve a comprehensive understanding of the experimental results. For calculation efficiency, combined fits were first performed for the diffuse neutron scattering patterns, INS spectra, and the saturation field without considering the SIA. Then the FMR resonance field was fitted by adding the SIA to the fitted model. Finally, combined fits were repeated for the diffuse neutron scattering patterns, INS spectra, and the saturation field with the SIA fixed at the fitted strengths. In the combined fits, the cost function, $\chi^2_\mathrm{A}$, for data A is defined as:
\begin{equation}
\renewcommand{\theequation}{S\arabic{equation}} 
\chi^2_\mathrm{A} = \frac{1}{N}\sum_i^N [(y_{i\mathrm{A}}^\mathrm{exp}-y_{i\mathrm{A}}^\mathrm{cal})/y_{i\mathrm{A}}^\mathrm{err}]^2,
\end{equation} 
where $y_{i\mathrm{A}}^\mathrm{exp}$ represents the $i$-th experiment data point for A, $y_{i\mathrm{A}}^\mathrm{cal}$ denotes the corresponding calculated value, and $y_{i\mathrm{A}}^\mathrm{err}$ is the experimental error. 
The total cost function is defined as: 
\begin{equation}
\renewcommand{\theequation}{S\arabic{equation}} 
\chi^2 = w_\mathrm{Diffuse}\chi_\mathrm{Diffuse}^2 + w_\mathrm{INS}\chi_\mathrm{INS}^2+w_\mathrm{Hsat}\chi_\mathrm{Hsat}^2,
\end{equation}
with the weighting factors $w_\mathrm{Diffuse}$, $w_\mathrm{INS}$, and $w_\mathrm{Hsat}$ fixed at 0.20, 0.55, and 0.25, respectively, to achieve comparable contributions to $\chi^2$ from each dataset. We confirm that slight adjustments of the weighting factors do not impact the fitting results within standard deviations. The following paragraphs summarize the fitting methodology and parameters used for each dataset.

The input for the diffuse scattering fits consists of absorption-corrected data in the ($h$, $k$, 1) plane, comprising 1,734 reciprocal space points sampled along the ($h$, 0, 0) and ($-k$, 2$k$, 0) directions in steps of 0.03~r.l.u. Notably, the diffuse scattering data in the ($h$, 0, $l$) plane, shown in Fig.~1(g) of the main text, was not included in the fitting procedure. The good agreement between this dataset and the calculated pattern shown in Fig.~1(i) of the main text therefore serves as an independent validation of the fitted model.
\begin{figure*}[t!]
    \centering
\renewcommand{\thefigure}{S\arabic{figure}}
    \includegraphics[width=0.8\linewidth]{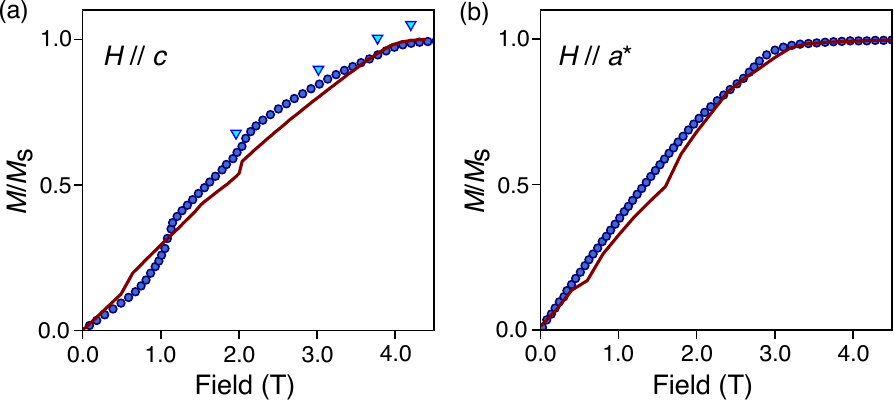}
    \hypertarget{anchor:Magnetization}{}
    \caption{Magnetization dependence of the field. (a,b) Experimental and calculated magnetization at $T=2$~K as a function of magnetic field along (a) $c$ and  (b) $a^*$ directions. The blue scatters are the experimental data as previously reported~\cite{hirschbergerSkyrmion2019s}, while the red lines are the calculated magnetization. Blue triangles in (a) indicate the magnetization at four characteristic field values that are considered in the multi-target fits.}
    \label{fig:Magnetization}
\end{figure*}
The input for the INS fits consists of the spectra shown in Fig.~2(a) of the main text, which is also reproduced in Fig.~\hyperlink{anchor:INS_fit}{\ref{fig:INS_fit}}, sampled with an energy step of 0.2~meV and a reciprocal space step of 0.075~r.l.u. along both $Q$ and $E$. To account for neutron absorption effects, the INS spectra were divided into six distinct regions, as indicated by the red boxes in Fig.~\hyperlink{anchor:INS_fit}{\ref{fig:INS_fit}}. Separate intensity scaling factors were introduced for each region to compensate for absorption-related variations.

The input for the magnetization fits was selected at four characteristic magnetic field values: $H = 2.0$, 3.0, 3.8, and 4.2~T, where the full magnetization data have been previously reported in Ref.~\cite{hirschbergerSkyrmion2019s}. Magnetization curves were calculated using classical Monte Carlo simulations with Langevin updates \cite{dahlbomSunnyjl2025,PhysRevB.106.235154} on a system of $N_\mathrm{cells} = 11 \times 11 \times 2$. Each simulation consisted of $1 \times 10^4$ thermalization sweeps, followed by $2 \times 10^4$ measurement sweeps, from which $2 \times 10^3$ decorrelated samples were extracted to ensure statistical reliability. The fitted magnetization curves are shown in Fig.~\hyperlink{anchor:Magnetization}{\ref{fig:Magnetization}}.

\begin{figure*}[b]
    \centering
\renewcommand{\thefigure}{S\arabic{figure}}
    \includegraphics[width=1\linewidth]{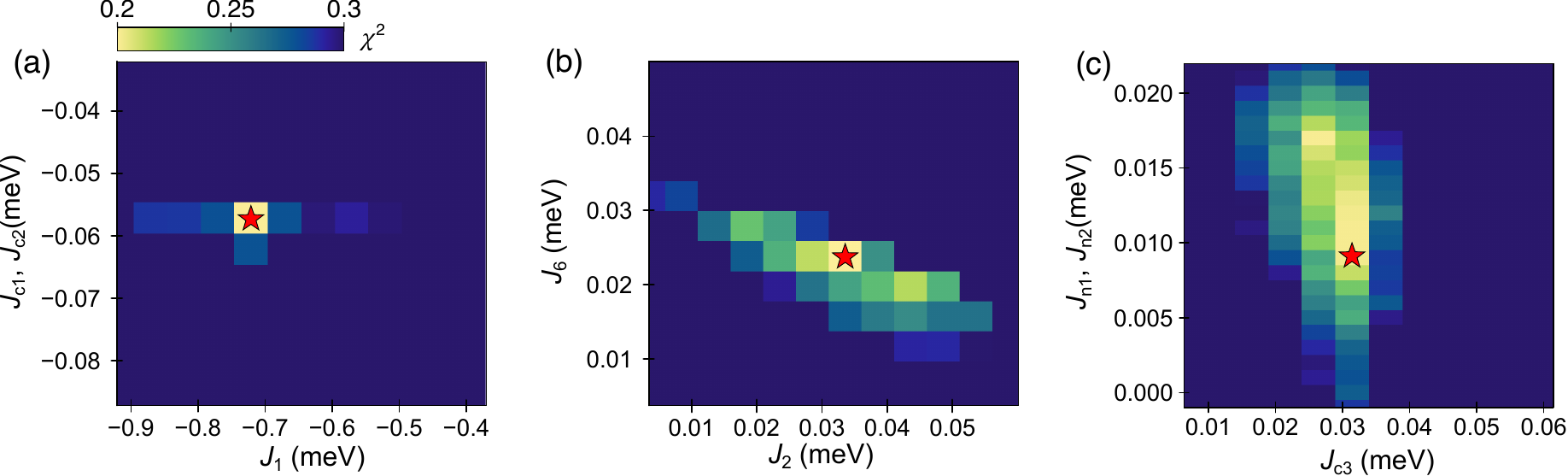}
    \hypertarget{anchor:chi2_mesh}{}
    \caption{
        Sensitivity maps for the fitted parameters. The goodness-of-fit parameter, $\chi^2$, as a function of the strengths for different coupling pairs, including $J_1$-$J_\mathrm{c1, c2}$ in (a), $J_2$-$J_\mathrm{6}$ in (b), and $J_\mathrm{c3}$-$J_\mathrm{n1,n2}$ in (c). In each panel, a red star marks the global minimum with the optimal parameter set as listed in the main text.
    } 
    \label{fig:chi_mesh}
\end{figure*}

For the $\chi^2$ minimization, we utilized the Marine Predators Algorithm (MPA) that was recently proposed in Ref.~\cite{FARAMARZI2020113377s}. MPA is a nature-inspired metaheuristic algorithm, which combines two types of stochastic moves described by the L\'evy flight and the Brownian motion. Although both types of moves are of the random walk style, the former searches the parameter space with small steps associated with occasional long jumps, while the latter covers the parameter space with more uniform steps. We used 96 search agents and set the fish aggregating device (FAD) to 0.5. The optimal parameters were obtained after reaching the 100th iteration.

For the fits to the FMR data, the strength of the SIA, $K_\mathrm{ab}$, was manually adjusted to reproduce the angular dependence of the resonance field reported in Ref.~\cite{hirschbergerLatticecommensurate2024s}. This separate fitting procedure is justified, as the resonance field is largely insensitive to the details of the isotropic exchange interactions, provided the system remains in the field-saturated ferromagnetic state. The fitted SIA strength is $K_\mathrm{ab} \sim 0.009$~meV, and the corresponding results are summarized in Fig.~\hyperlink{anchor:ESR}{\ref{fig:ESR}}. In addition to the easy-plane SIA, the dipole-dipole interactions are found to contribute significantly to the observed anisotropy.

\begin{figure*}[t]
    \centering
\renewcommand{\thefigure}{S\arabic{figure}}
    \includegraphics[width=0.7\linewidth]{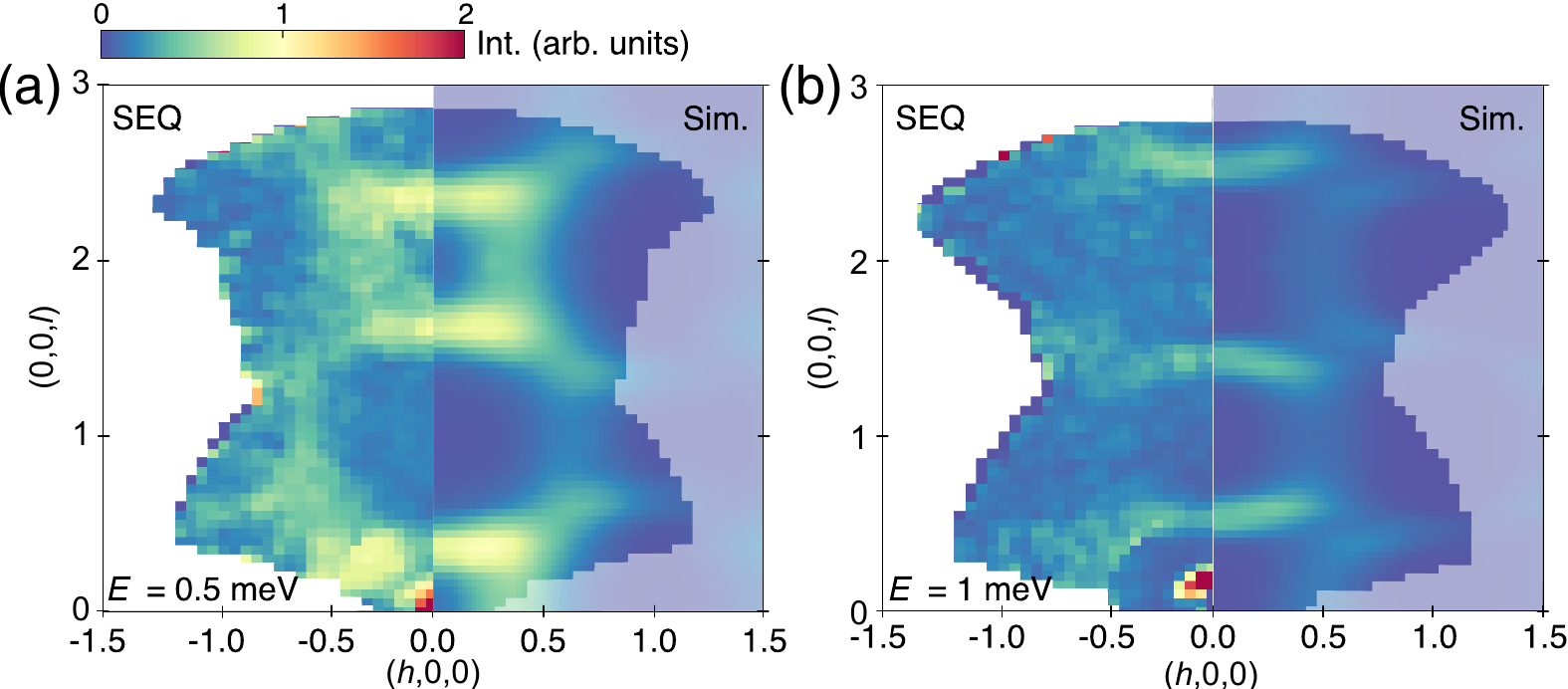}
    \hypertarget{anchor:constant_energy_slice1}{}
    \caption{Comparison to the INS spectra collected with a lower incident neutron energy. Constant-energy slices of experimental and calculated INS spectra in the ($h$, 0, $l$) plane, measured with an incident energy $E_\mathrm{i}$ of 8~meV, at energy transfers of $E = 0.5$~meV (a) and 1.0~meV (b). The integration widths are $\Delta E = \pm$0.2~meV along the energy axis and $\Delta k = \pm$0.2~r.l.u. along the $(-k, 2k, 0)$ axis. In each panel, the experimental and calculated results are displayed on the left and right halves, respectively.} 
    \label{fig:INS_cut}
\end{figure*}

Figure~\hyperlink{anchor:chi_mesh}{\ref{fig:chi_mesh}} presents the sensitivity maps of the fitted coupling strengths. For clarity, the six independent coupling parameters $J_{126}$-$J_{\rm{c123}}$-$J_{\rm{n12}}$ are grouped into three pairs, including the $J_1$-$J_\mathrm{c1,c2}$, $J_2$-$J_\mathrm{6}$, and $J_\mathrm{c3}$-$J_\mathrm{n1,n2}$ pairs, and the red star in each panel marks the optimal parameter set as listed in Table~1 of the main text. Analysis of the sensitivity maps reveals distinct impacts of the coupling parameters. For example, as revealed in Fig.~\hyperlink{anchor:chi_mesh}{\ref{fig:chi_mesh}(a)}, $\chi^2$ is markedly less sensitive to variations in $J_1$. This observation can be explained by the fact that ferromagnetic intra-trimer interactions, as long as they dominate, only marginally impact the diffuse scattering pattern and the magnon band at lower energies. Furthermore, the sensitivity map for the pair of $J_2$ and $J_6$ interactions in Fig.~\hyperlink{anchor:chi_mesh}{\ref{fig:chi_mesh}(b)} reveals a correlation between these two parameters: The $\chi^2$ minima follow a line of $2\Delta J_{2}-\Delta J_{6}$ around the optimal parameter set. As discussed in the main text, this correlation arises from the similar impacts of $J_2$ and $2J_6$ on the effective honeycomb lattice as both couplings contribute to the intra-sublattice interactions.

As further evidence for the accuracy of our fitted model, Fig.~\hyperlink{anchor:INS_cut}{\ref{fig:INS_cut}} compares the calculated INS spectra to the high-resolution experimental data measured with a lower incident neutron energy of $E_\mathrm i=8$~meV on SEQUOIA. The constant-energy slices are integrated at an energy transfer of 0.5 (\hyperlink{anchor:INS_cut}{a}) and 1~meV (\hyperlink{anchor:INS_cut}{b}), with an integration width of $\Delta E = \pm 0.2$~meV. Close agreement between the experimental and simulated data further validates our microscopic model. 

\section{Calculation of the skyrmion number $N_\mathrm{sk}$}
In our classical Monte Carlo simulations, we compute the skyrmion number, $N_\mathrm{sk}$, to characterize the topological phases. This quantity is defined as the sum of the solid angles subtended by the spins on all triangular plaquettes within a magnetic unit cell:
\begin{equation}
\renewcommand{\theequation}{S\arabic{equation}} 
N_\mathrm{sk} = \frac{1}{4\pi} \sum_{i,j,k \in \triangle} \Omega_{ijk}.
\end{equation} 

Here, $\Omega_{ijk}$ is the solid angle over a triangle formed by sites $i$, $j$, and $k$, 

\begin{equation}
\renewcommand{\theequation}{S\arabic{equation}} 
\Omega_{ijk} = 2\tan^{-1} \frac{\mathbf{S}_i\cdot(\mathbf{S}_j\times \mathbf{S}_k)}{|\mathbf{S}_i||\mathbf{S}_j||\mathbf{S}_k|+\mathbf{S}_i \cdot \mathbf{S}_j |\mathbf{S}_k| + \mathbf{S}_k \cdot \mathbf{S}_i |\mathbf{S}_j| + \mathbf{S}_k \cdot \mathbf{S}_j |\mathbf{S}_i|}\textrm{ ,}
\label{eq:solidangle}
\end{equation} 
where $\mathbf{S}_i\cdot(\mathbf{S}_j\times \mathbf{S}_k)$ is the scalar spin chirality on the triangle $ijk$.

\begin{figure*}[t]
    \centering
\renewcommand{\thefigure}{S\arabic{figure}}
    \includegraphics[width=0.35\linewidth]{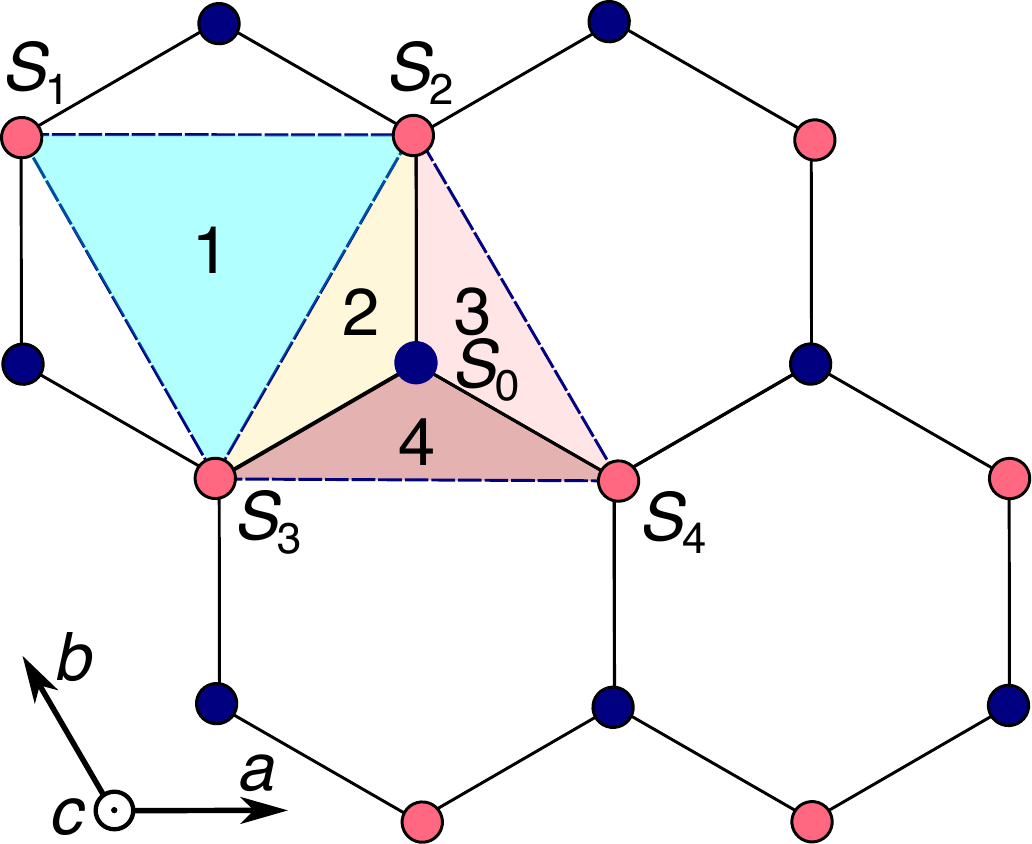}
    \hypertarget{anchor:N_sk_cal}{}
    \caption{Schematic diagram of triangulation for the skyrmion number calculation. The blue and red sites represent the A and B sublattices, respectively, of the AB-stacked triangular lattice. Each unit cell is divided into four triangles denoted as $T_{ijk}$, where the subscripts $ijk$ represent the indices at the three corners of each triangle.} 
    \label{fig:Nsk_cal} 
\end{figure*}

To implement this calculation, the original breathing kagome lattice and the effective triangular lattice are partitioned into triangular plaquettes. Figure~\hyperlink{anchor:Nsk_cal}{\ref{fig:Nsk_cal}} illustrates this triangulation on the effective triangular lattice: the unit cell formed by sites 1-4 of the B sublattice is divided into four triangular plaquettes, including $\triangle_{132}$ from sites on the B sublattice and $\triangle_{023}$, $\triangle_{042}$, and $\triangle_{034}$ around site $S_0$ of the A sublattice.  All triangles are consistently traversed in the anticlockwise direction to preserve the orientation required for chirality calculations.

\begin{figure*}[b]
    \centering
\renewcommand{\thefigure}{S\arabic{figure}}
    \includegraphics[width=0.8\linewidth]{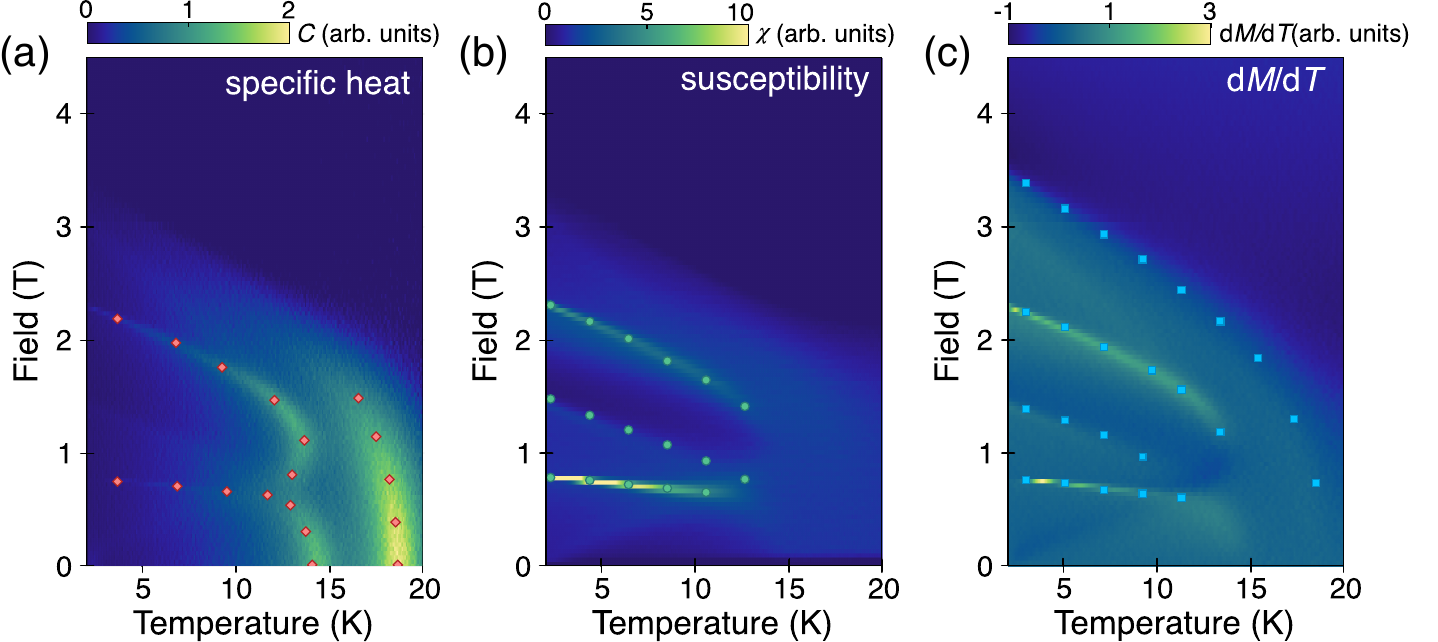}
    \hypertarget{anchor:Heat_chi}{}
    \caption{Location of the transition points in the phase diagram. Calculated specific heat (a), magnetic susceptibility (b) and derivative of magnetization with respect to temperature (c) as a function of temperature and magnetic field. Red squares and green circles mark the phase boundaries in heat capacity and magnetic susceptibility, respectively. The same markers are shown in Fig.~3(a) of the main text.}
    \label{fig:heat_chi}
\end{figure*}

\begin{figure*}[t]
    \centering
\renewcommand{\thefigure}{S\arabic{figure}}
    \includegraphics[width=0.9\linewidth]{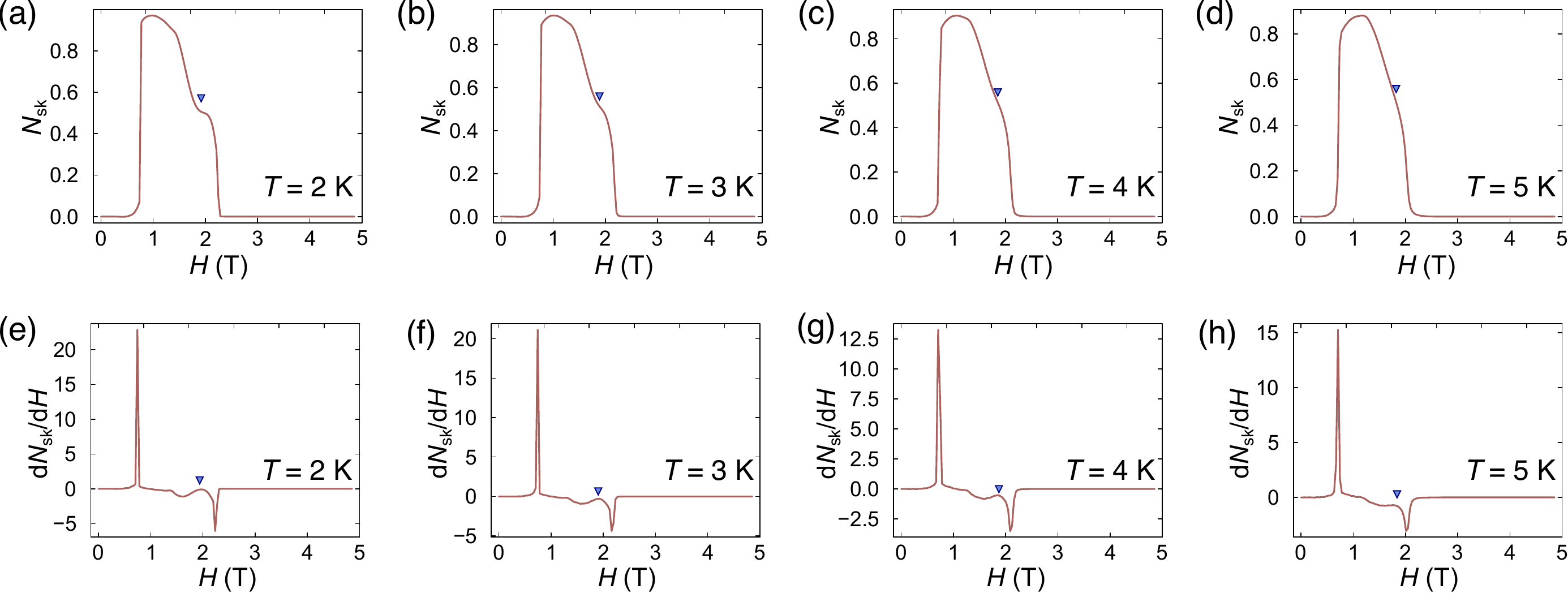}
    \hypertarget{anchor:N_sk_c}{}
    \caption{Transition between the skyrmion and inverted meron phases. Skyrmion number, $N_\mathrm{sk}$, and its derivative, $dN_\mathrm{sk}/dH$, calculated at $T = 2$~K (a and e), 3~K (b and f), 4~K (c and g), and 5~K (d and h). The blue triangles indicate the transition point at which the inverted meron phase with half skyrmion number appears. The same markers are shown in Fig.~3(a) of the main text.}
    \label{fig:Nsk_c}
\end{figure*}

\section{Phase boundary determination in the phase diagram}
The high-resolution phase diagram shown in Fig.~3(a) of the main text was calculated on an effective triangular lattice of $N_\mathrm{cells}=12\times 12\times 4$ using parallel tempering with 288 replicas at different temperatures. Simulations consisted of an initial $2\times10^5$ sweeps for thermalization, followed by $5\times10^5$ sweeps for measurement. Replica exchange attempts were performed every five sweeps. The phase boundaries shown in the figure were determined from the temperature and field evolution of the specific heat, magnetic susceptibility, and skyrmion number. Figure~\hyperlink{anchor:heat_chi}{\ref{fig:heat_chi}} presents the colormaps for the specific heat, magnetic susceptibility, and derivative of the magnetization with respect to temperature, in which the phase transitions were marked by red diamonds, green dots, and blue squares, respectively. The same markers are presented in Fig.~3(a) of the main text. Notably, weak transitions that are barely discernible in specific heat become more pronounced in magnetic susceptibility. This includes the transition from the stripe phase to the meron-antimeron phase and the transition from the meron-antimeron phase to the skyrmion phase. 
 
The boundary between the skyrmion and inverted meron phases is determined from the field evolution of the skyrmion number. Figures~\hyperlink{anchor:Nsk_c}{\ref{fig:Nsk_c}(a-d)} present the skyrmion number as a function of field calculated at different temperatures. Their corresponding derivatives, $dN_\mathrm{sk}/dH$, are shown in Figs.~\hyperlink{anchor:Nsk_c}{\ref{fig:Nsk_c}(e-h)}. At $T=2$~K, the inverted meron phase can be identified by the $N_\mathrm{sk}=1/2$ plateau shown in Fig.~\hyperlink{anchor:Nsk_c}{\ref{fig:Nsk_c}(a)}. This plateau becomes gradually blurred at elevated temperatures, leading to a kink in $N_\mathrm{sk}(H)$ at $T=5$~K. The boundary between the skyrmion and inverted meron phases can be determined using the local maximum in $dN_\mathrm{sk}/dH$, which is marked by a blue triangle in each panel.

\begin{figure*}[t]
    \centering
    \renewcommand{\thefigure}{S\arabic{figure}}
    \includegraphics[width=0.8\linewidth]{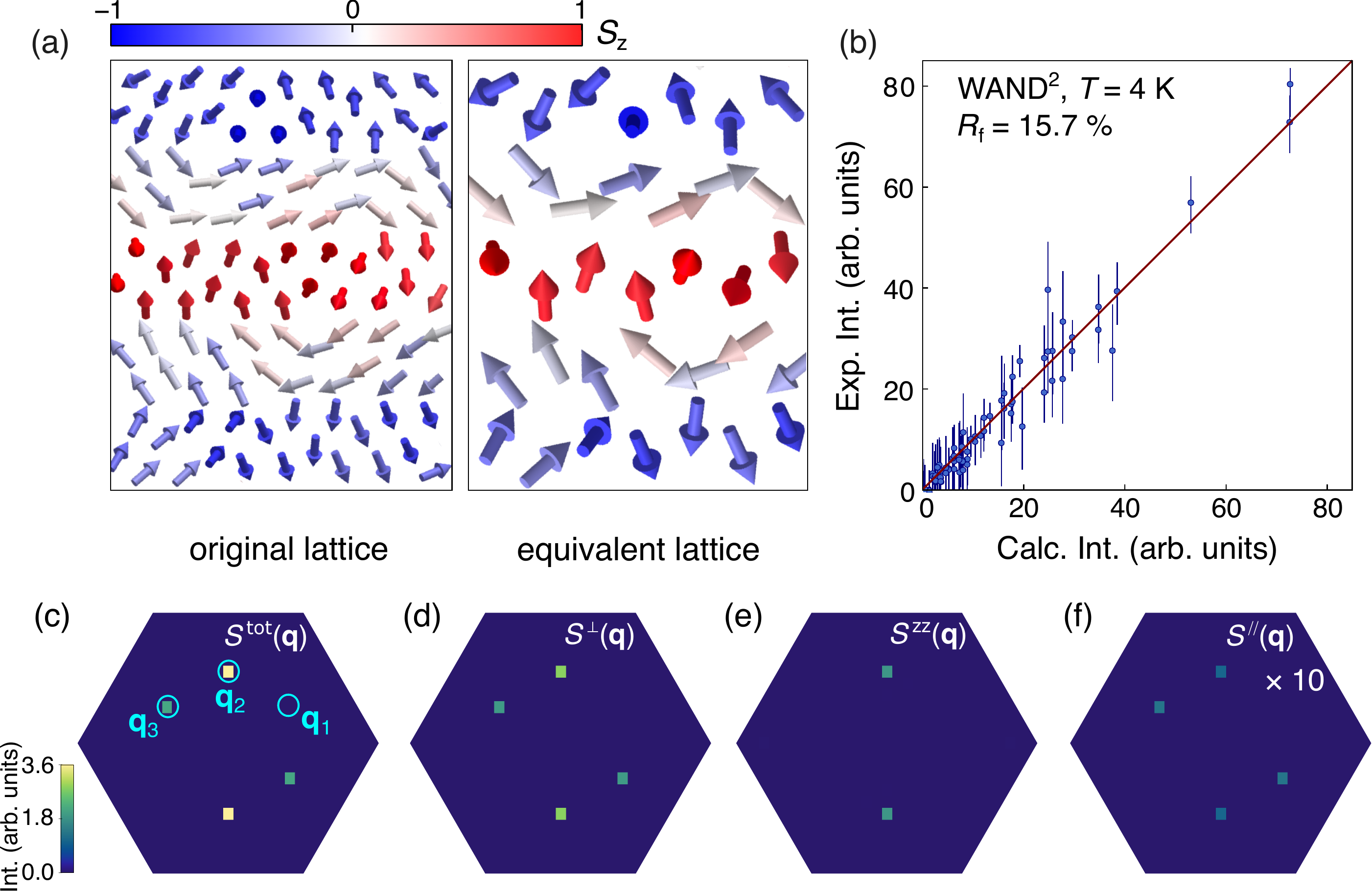} 
    \hypertarget{anchor:DoubleQ}{}
    \caption{Magnetic ground state of Gd$_3$Ru$_4$Al$_{12}$. (a) Simulated magnetic ground state of Gd$_3$Ru$_4$Al$_{12}$ viewed along the $c$ axis on the original breathing kagome lattice (left) and on the equivalent triangular lattice of effective spins (right). Pseudocolor corresponds to the length of the spin component along the $c$ axis. (b) Comparison of the experimental and calculated intensity of the magnetic Bragg peaks with a goodness-of-fit parameter of $R_\mathrm{f}=15.7~\%$. The experimental neutron diffraction dataset were collected on WAND$^2$ at $T =4$~K. (c) The total magnetic structure factors, $S^{\mathrm{tot}}(\mathbf{q})$, for the double-\textbf{Q} stripe phase calculated on the original lattice. Blue circles in panel (c) mark the three magnetic propagation vectors, $\mathbf{q}_1$, $\mathbf{q}_2$, and $\mathbf{q}_3$. The total magnetic structure factors are decomposed into the $S^\bot(\mathbf{q})$ (panel d), $S^\mathrm{zz}(\mathbf{q})$ (panel e), and $S^{\mathrel{/\negmedspace/}}(\mathbf{q})$ (panel f) components, which represent the components that are perpendicular to both $\mathbf{q}$ and the $c$ axis, along the $c$ axis, and along the $\mathbf{q}$ vectors, respectively.}  
    \label{fig:doubleQ}
\end{figure*}  
\section{The Double-Q ground state}
Figure~\hyperlink{anchor:doubleQ}{\ref{fig:doubleQ}(a)} presents the magnetic structure of the stripe order in the ground state, with the same spin texture being described in the left panel for the original breathing kagome lattice and the right panel for the effective triangular lattice. Figure~\hyperlink{anchor:doubleQ}{\ref{fig:doubleQ}(b)} compares the experimental intensity of the neutron diffraction dataset collected on WAND$^2$ at $T = 4$~K and the calculated intensity for the stripe order, which yields excellent agreement with a goodness-of-fit parameter $R_\mathrm{f}=15.7~\%$.

The double-${\mathbf Q}$ character of the stripe order is evidenced by the equal-time spin structure factor $\mathcal{S}^{\mathrm{tot}}(\mathbf{q})$ as shown in Fig.~\hyperlink{anchor:doubleQ}{\ref{fig:doubleQ}(c)}, which reveals the coexistence of two arms, $\mathbf{q}_2$ and $\mathbf{q}_3$, with unequal intensities. As shown in Figs.~\hyperlink{anchor:doubleQ}{\ref{fig:doubleQ}(d-f)}, the total spin structure factor can be further decomposed into the $\mathcal{S}^\bot (\mathbf{q})$, $\mathcal{S}^{\mathrel{/\negmedspace/}}(\mathbf{q})$, and $\mathcal{S}^\mathrm{zz} (\mathbf{q})$ components that are defined as
\renewcommand{\theequation}{S\arabic{equation}} 
 \begin{align}
    \mathcal{S}^{\bot}(\mathbf{q}) &= \frac{1}{q^2}\mathbf{q}_{\bot}^\mathrm{T} \mathcal{S} (\mathbf{q})\mathbf{q}_{\bot}\textrm{\qquad for the component perpendicular to both $\mathbf{q}$ and the $c$ axis,}
    \label{Eq:S_perp} \\
    \mathcal{S}^{\mathrel{/\negmedspace/}}(\mathbf{q}) &= \frac{1}{q^2}\mathbf{q}^\mathrm{T} \mathcal{S}(\mathbf{q})\mathbf{q}\textrm{\qquad\;\, for the component parallel with $\mathbf{q}$,}
    \label{Eq:S_para} \\
    \mathcal{S}^\mathrm{zz}(\mathbf{q}) &= \mathbf{\hat{n}_z^\mathrm{T}} \mathcal{S}(\mathbf{q})\mathbf{\hat{n}_z}\textrm{\qquad\quad\ for the component parallel with the $c$ axis.}
    \label{Eq:S_zz}
\end{align}
\begin{figure*}[t]
    \centering
\renewcommand{\thefigure}{S\arabic{figure}}
    \includegraphics[width=0.95\linewidth]{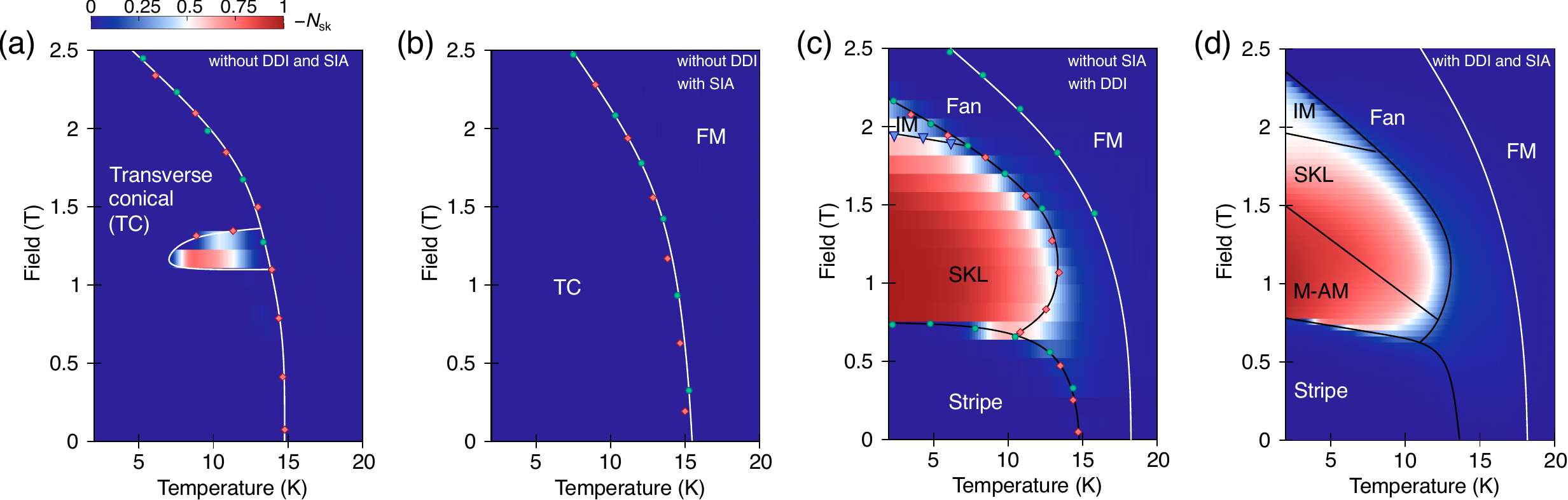}
    \hypertarget{anchor:Dip_aniso}{}
    \caption{Impacts of the dipole-dipole interactions and the single-ion anisotropy. Comparison of the phase diagrams calculated for the fitted exchange coupling strengths under different conditions: (a) without the DDI or SIA, (b) with SIA but without DDI, (c) with DDI but without SIA, and (d) with both DDI and SIA as that shown in Fig.~3(a) of the main text. If present, the strengths of the DDI and SIA are the same as that employed in the full model. Pseudocolor corresponds to the negative skyrmion number, $-N_\mathrm{sk}$.}
    \label{fig:Dip_aniso}
\end{figure*}
In these expressions, $\mathbf{q}_\bot$ is obtained by a counterclockwise $\pi/2$ rotation of the $\mathbf{q}$ vector around the $c$ axis, and $\mathbf{n_z}$ is the unit vector along the $c$ axis. The decomposition of the structure factor reveals that the $\mathcal{S}^{\mathrel{/\negmedspace/}}(\mathbf{q})$ component is much weaker compared to the $\mathcal{S}^{\bot}(\mathbf{q})$ and $\mathcal{S}^\mathrm{zz}(\mathbf{q})$ components, which is consistent with the polarization analysis of the RXD experiments~\cite{hirschbergerSkyrmion2019s}. Comparison of the $\mathcal{S}^{\bot}(\mathbf{q})$ component in Fig.~\hyperlink{anchor:doubleQ}{\ref{fig:doubleQ}(d)} and the $\mathcal{S}^\mathrm{zz}(\mathbf{q})$ component in Fig.~\hyperlink{anchor:doubleQ}{\ref{fig:doubleQ}(e)} reveals that the main arm, $\mathbf{q}_2$, forms a helical order with reduced ordering moments along the $c$ axis, while the secondary arm, $\mathbf{q}_3$, forms a sinusoidally modulated spin density wave order with spins along the $\mathbf{q}_\bot$ direction. This decomposition also agrees with the polarization analysis of the RXD experiments~\cite{hirschbergerSkyrmion2019s}. Typically, easy-plane anisotropy favors a cycloidal spin structure within the $ab$ plane, rather than a helix. This stripe ground state structure in Gd$_3$Ru$_4$Al$_{12}$ emerges from the interplay of easy-plane anisotropy, dipole-dipole interaction, and magnetic frustration. 

\section{Origin of the topological phases and impacts of the SIA and DDI}
To elucidate the origin of the topological phases in Gd$_3$Ru$_4$Al$_{12}$, we compared phase diagrams calculated using the fitted $J_{126}$--$J_{\rm{c123}}$--$J_{\rm{n12}}$ model on a $12\times12\times4$ supercell of the effective triangular lattice, with and without the inclusion of SIA and DDI. As shown in Fig.~\hyperlink{anchor:Dip_aniso}{\ref{fig:Dip_aniso}(a)}, even in the absence of SIA and DDI, a narrow region of skyrmion phase is already present, confirming that magnetic frustration is a key ingredient~\cite{Okubo2012s}.  

Introducing an easy-plane SIA with $K_\mathrm{ab} = 0.009$~meV into the model suppresses the skyrmion phase entirely, as shown in Fig.~\hyperlink{anchor:Dip_aniso}{\ref{fig:Dip_aniso}(b)}. In contrast, the inclusion of DDI stabilizes the skyrmion phase and significantly enlarges its extent in the phase diagram, as illustrated in Fig.~\hyperlink{anchor:Dip_aniso}{\ref{fig:Dip_aniso}(c)}. When both SIA and DDI are incorporated, a meron--antimeron phase emerges within the topological regime, as shown in Fig.~\hyperlink{anchor:Dip_aniso}{\ref{fig:Dip_aniso}(d)} and Fig.~3(a) of the main text.  
This comparison highlights the crucial role of the easy-plane SIA in stabilizing the meron--antimeron phase.

\begin{figure*}[b]
    \centering
\renewcommand{\thefigure}{S\arabic{figure}}
    \includegraphics[width=0.9\linewidth]{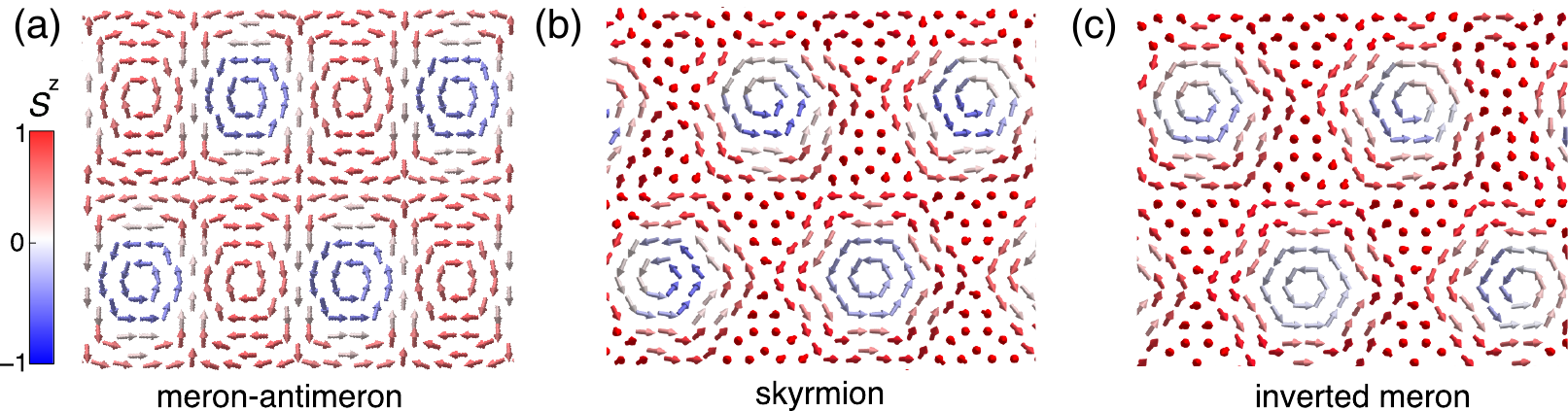}
    \hypertarget{anchor:ori_structure}{}
    \caption{Simulated spin textures on the original breathing kagome lattice. Simulated magnetic structures in the (a) meron-antimeron, (b) skyrmion, and  (c) inverted meron phases viewed along the $c$ axis. Simulations were performed on the orignal breathing kagome lattice using the fitted parameter set as listed in the main text. Pseudocolor of the spins corresponds to the length of spin component along the $c$ axis.}
    \label{fig:ori_structure}
\end{figure*}

\begin{figure*}[t]
    \centering
\renewcommand{\thefigure}{S\arabic{figure}}
    \includegraphics[width=0.8\linewidth]{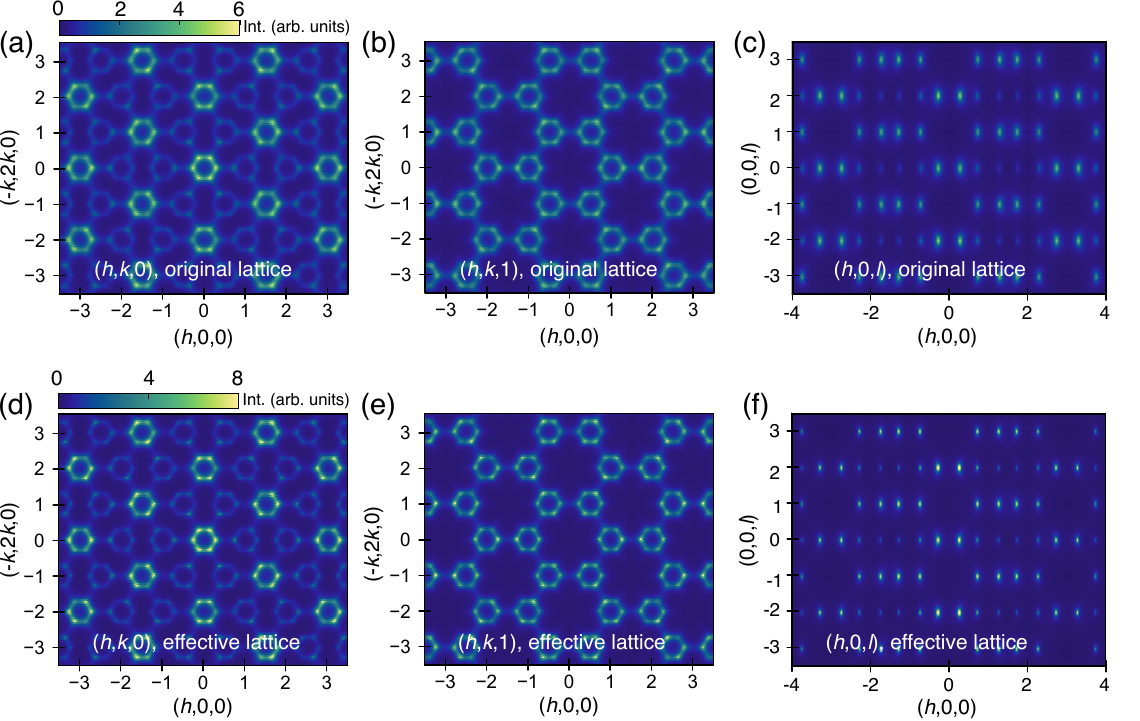}
    \hypertarget{anchor:diffuse}{}
    \caption{Comparison of the short-range spin correlations on the original breathing kagome lattice and the effective triangular lattice. Diffuse scattering patterns in the ($h$, $k$, 0), ($h$, $k$, 1) and ($h$, 0, $l$) planes for the fitted model on the effective AB-stacked triangular lattice (a-c) and the original AB-stacked breathing kagome lattice (d-f).}
    \label{fig:diffuse}
\end{figure*}

\section{Simulations on the original and effective lattices}

In the original breathing kagome lattice of Gd$_3$Ru$_4$Al$_{12}$, the strong intra-trimer coupling $J_1$ presents significant challenges for Monte Carlo simulations, as the acceptance rate for flipping a single spin within a trimer becomes vanishingly small at temperatures $T \ll J_1S^2$. This difficulty is particularly pronounced when calculating the wavevector shift, since the magnetic propagation vector can easily become trapped in a local minimum, making the final state highly sensitive to the initial magnetic configuration.  
To circumvent this issue, we employed the equivalent model on the effective triangular lattice, which enabled the calculation of a high-resolution phase diagram, as shown in Fig.~3(a) of the main text.

To justify our approach using the effective triangular-lattice model, we calculated the magnetic ground states on the original breathing kagome lattice using the conjugate gradient method. This method is well-known for its efficiency in locating the state with minimal energy at zero temperature. Figure~\hyperlink{anchor:ori_structure}{\ref{fig:ori_structure}} displays the calculated structures under magnetic fields of 1.0~T (\hyperlink{anchor:ori_structure}{a}), 1.9~T (\hyperlink{anchor:ori_structure}{b}), and 2.4~T (\hyperlink{anchor:ori_structure}{c}). To ensure that the global minimum was found, each calculation was repeated 100 times with random initial spin configurations, and the state with the lowest energy was selected. Due to the wavevector shift, the calculations for the meron-antimeron phase shown in Fig.~\hyperlink{anchor:ori_structure}{\ref{fig:ori_structure}(a)} were performed on a $12 \times 12 \times 4$ supercell. For the skyrmion (Fig.~\hyperlink{anchor:ori_structure}{\ref{fig:ori_structure}(b)}) and inverted meron (Fig.~\hyperlink{anchor:ori_structure}{\ref{fig:ori_structure}(c)}) phases, the calculations were carried out on a $11 \times 11 \times 4$ supercell to minimize boundary effects.

Further justification of the effective lattice approach can be drawn from the spiral spin-liquid phase. Figure~\hyperlink{anchor:diffuse}{\ref{fig:diffuse}} compares the diffuse neutron scattering patterns on the original lattice and the effective triangular lattice calculated through the classical Monte Carlo simulations with Langevin updates. Calculations were performed at $T=2$~K on a $22\times22\times2$ supercell for both models. To enable direct comparison,  phase factors of the three spins belonging to the same $J_1$ trimer on the original lattice was uniformly assigned to $\exp(i\mathbf{k}\cdot\mathbf{r}_0)$, with $\mathbf{r}_0$ representing the center position of the trimer \cite{gao_hierarchical_2021}. The resulting patterns are nearly identical on the original and effective lattices, with marginal differences arising from the finite strength of $J_1$ on the original lattice. This similarity in the short-range spin correlations further validates the equivalent triangular lattice as a computationally efficient approximation of the original breathing kagome lattice. 
\section{Polarization analysis in resonant x-ray diffraction}
\begin{figure*}[t]
    \centering
    \renewcommand{\thefigure}{S\arabic{figure}}
    \includegraphics[width=0.4\linewidth]{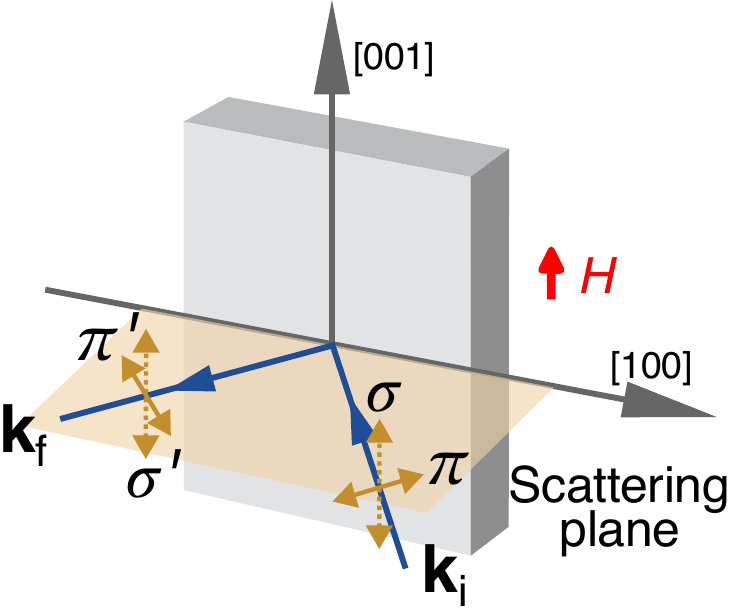} 
    \hypertarget{anchor:RXS}{}
    \caption{A schematic of the resonant x-ray diffraction experiment featuring polarization analysis. The $\pi$ and $\sigma$ notations indicate the x-ray polarization parallel and perpendicular, respectively, to the horizontal scattering plane (highlighted in yellow).}  
    \label{fig:RXS}
\end{figure*} 

As explained in the main text, polarization analysis of the previously reported resonant x-ray diffraction data~\cite{hirschbergerLatticecommensurate2024s} provides further evidence for the commensurate meron--antimeron (M-AM) spin texture. Figure~\hyperlink{anchor:RXS}{\ref{fig:RXS}} describes the scattering geometry of the x-ray experiment, where the ($h$, $k$, 0) plane was aligned as the horizontal scattering plane. The incident x-ray was linearly polarized parallel to the scattering plane ($\pi$). The polarization components $\sigma'$ (perpendicular to the scattering plane) and $\pi'$ (parallel to the scattering plane) of the scattered x-ray were measured separately using a PG analyzer. The scattering amplitudes can be expressed as:
\begin{align}
    f_{\text{res}} &\propto (\mathbf{e}_\mathrm i\times \mathbf{e}_\mathrm f') \cdot \mathbf{m}(\mathbf{Q}) \textrm{,}
\end{align}
where $\mathbf{e}_\mathrm i$ and $\mathbf{e}_ \mathrm f$ are the unit incident and scattered wavevectors, respectively, and $\mathbf{m}(\mathbf{q})$ is the magnetic structural factor. For the $\pi \to \sigma'$ channel, the scattering amplitude is proportional to the projection of the magnetic moment onto the incident wavevector $\mathbf e_\mathrm i$, thus probing the magnetic components within the scattering plane:
\begin{equation}
    f_{\text{res}}^{\pi \to \sigma'} \propto \mathbf e_\mathrm{i} \cdot \mathbf m(\mathbf Q)\textrm{.}
\end{equation}
In contrast, for $\pi \to \pi'$ channel, the cross product $(\mathbf{k}_\mathrm i \times \mathbf{k}_\mathrm f)$ is  along the $z$ direction that is parallel with the $\bm{c}$ axis, its dot product with $\mathbf{m}(\mathbf{Q})$ isolates the $m_\mathrm z(\mathbf{Q})$ component. The magnitude of the cross product is $|\mathbf{e}_\mathrm i \times \mathbf{e}_\mathrm f| = \sin(2\theta)$, where $2\theta$ is the scattering angle between $\mathbf{e}_\mathrm i$ and $\mathbf{e}_\mathrm f$. Therefore, the amplitude for the $\pi \to \pi'$ channel is proportional to the out-of-plane magnetic component:
\begin{equation}
    f_{\text{res}}^{\pi \to \pi'} \propto m_\mathrm z(\mathbf{Q}) \sin(2\theta)\textrm{.}
\end{equation}
The measured scattering intensity $I$ in each channel is proportional to the squared magnitude of the corresponding scattering amplitude, $I \propto |f_{\text{res}}|^2$. Consequently, the intensities in the two analyzed channels are directly sensitive to different components of the magnetic moment vector:
\begin{align}
    I^{\pi \to \sigma'} &\propto |\mathbf{e}_\mathrm i \cdot \mathbf{m}(\mathbf{Q})|^2 \textrm{,}\\
    I^{\pi \to \pi'} &\propto |m_\mathrm z(\mathbf{Q}) \sin(2\theta)|^2 \textrm{.}
\end{align}
The ratio $R(\mathbf{Q})$ between the intensity in these two channels, after normalized by the $\sin^2(2\theta)$ geometry factor, can be expressed as
\begin{align}
R(\mathbf{Q}) \sin^2(2\theta)=\frac{I_{\pi\to\sigma'}}{I_{\pi\to\pi'}}\sin^2(2\theta) = \frac{|\mathbf{e}_\mathrm i \cdot \mathbf m(\mathbf{Q})|^2}{|m_\mathrm z(\mathbf Q)|^2} \textrm{.}
\end{align}
Following this definition, we can directly compare the experimental and calculated $R(\mathbf{Q})\sin^2(2\theta)$ values for each magnetic Bragg peaks, where the experimental values have been tabulated in the Supplementary Table 1 of Ref.~\cite{hirschbergerLatticecommensurate2024s}. For a more detailed analysis, particularly for helical or cycloidal spin structures, we can further decompose the in-plane magnetic structure factor $\mathbf m _\mathrm {xy} (\mathbf Q)$ into its magnitude $|\mathbf m _\mathrm {xy}(\mathbf Q)|$ and its directional unit vector $\mathbf e_\perp (\mathbf Q)$. The numerator then becomes $|\mathbf e_\mathrm i \cdot \mathbf m_\mathrm{xy}(\mathbf Q)|^2= S^\perp(\mathbf{Q})\cdot [\mathbf e_\mathrm i\cdot \mathbf e_\perp (\mathbf Q)]^2+S^\parallel(\mathbf{Q})\cdot \mathbf [\mathbf e_\mathrm i \cdot \mathbf e_\parallel (\mathbf Q)]^2$. This allows to rewrite the ratio equation as: 
\begin{equation}
R(\mathbf{Q}) \sin^2(2\theta) =\frac{S^\perp(\mathbf{Q})}{S^\mathrm{zz}(\mathbf{Q})}\cdot [\mathbf{e}_\mathrm i \cdot \mathbf{e}_\perp(\mathbf{Q})]^2+ \frac{S^\parallel(\mathbf{Q})}{S^\mathrm{zz}(\mathbf{Q})}\cdot [\mathbf{e}_\mathrm i \cdot \mathbf{e}_\parallel(\mathbf{Q})]^2 \textrm{.} 
\end{equation}
Therefore, by plotting the experimentally measured quantity $R(\mathbf Q)\sin^2(2\theta)$ against the calculated geometrical factor $[\mathbf e_\mathrm i \cdot \mathbf e_\perp (\mathbf Q)]^2$, which can be varied by rotating the sample azimuthally, the slope of the resulting line directly yields the ratio of the in-plane to out-of-plane magnetic structure factors, while the intercept corresponds to $\frac{S^\parallel(\mathbf{Q})}{S^\mathrm{zz}(\mathbf{Q})}\cdot [\mathbf{e}_\mathrm i \cdot \mathbf{e}_\parallel(\mathbf{Q})]^2$. This method provides a powerful way to quantitatively determine the nature of the spin texture.
For the calculations, we considered the M-AM and skyrmion lattices (SKL) on a 12$\times$12$\times$2 supercell of the original breathing kagome lattice. The spin configurations were first initialized on the corresponding structures, and then optimized through local energy minimization in different magnetic field. The magnetic structure factor, $\mathbf m(\mathbf Q)$, is calculated for each magnetic Bragg peaks. For the M-AM lattice that is characterized by dominant double-$\mathbf q$ components, equal domain population was assumed when calculating the scattering intensity. Comparison between the experimental and calculated $R(\mathbf{Q})$ values are presented in Fig.~4 of the main text.

\begin{figure*}[t]
    \centering
\renewcommand{\thefigure}{S\arabic{figure}}
    \includegraphics[width=0.8\linewidth]{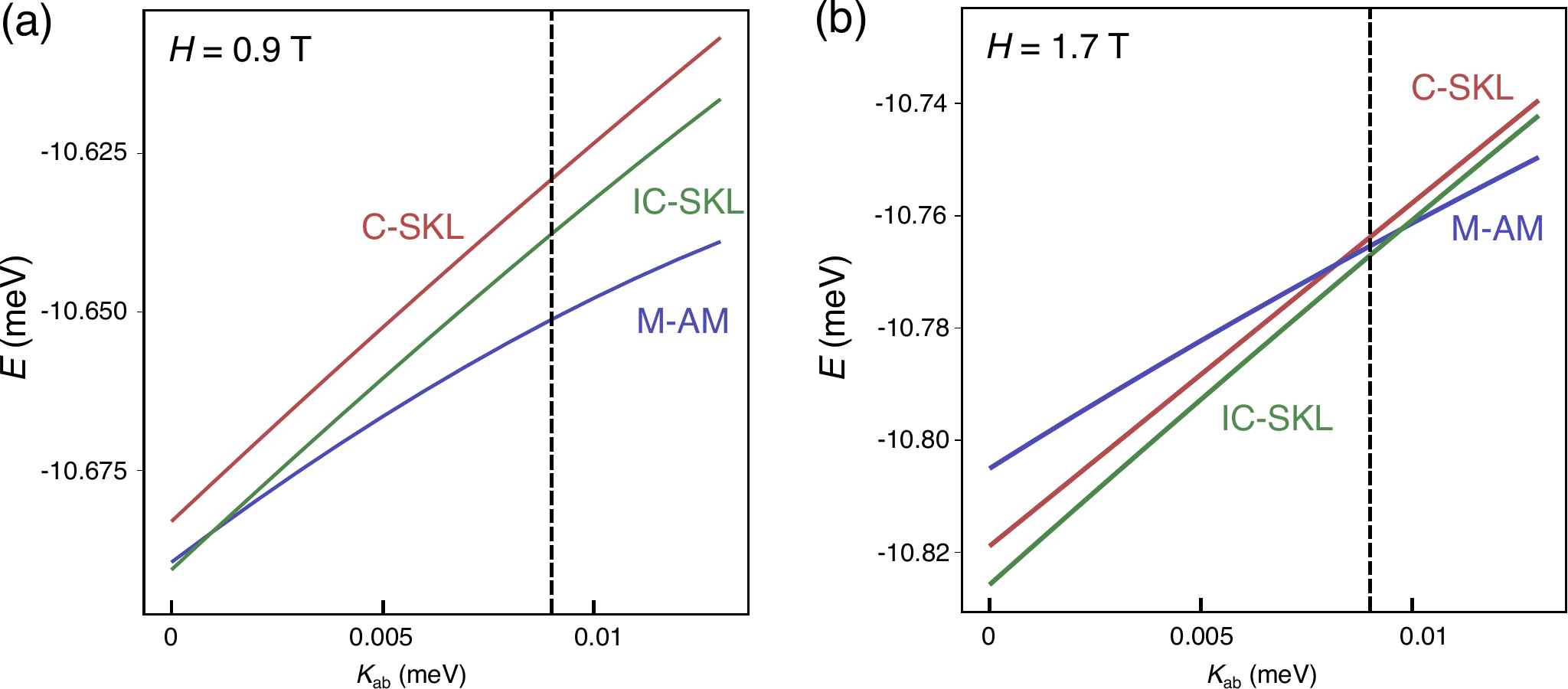}
    \hypertarget{anchor:Ecomp}{}
    \caption{Energy per site as a function of easy-plane anisotropy ($K_\mathrm{ab}$) for three competing magnetic phases: commensurate skyrmion lattice (C-SKL), incommensurate skyrmion lattice (IC-SKL), and M-AM. Calculations are presented for (a) a low magnetic field ($H$ = 0.9~T) and (b) a high magnetic field ($H$ = 1.7~T). The energy of the IC-SKL state is consistently lower than that of the C-SKL state, demonstrating that the relaxation of the wave vector from a fixed commensurate value is an energetically favorable process that stabilizes the system.}
    \label{fig:E_comp}
\end{figure*}

\section{M-AM lattice versus the commensurate SKL}
The previous study~\cite{hirschbergerLatticecommensurate2024s}, which is based on a phenomenological model, proposed the existence of a commensurate SKL lattice (C-SKL) in the lower field regime of the topological phase. In contrast, our microscopic modeling reveals the emergence of a M-AM spin texture in the same commensurate regime. The microscopic model established in our work enables a quantitative comparison for the stability of the commensurate M-AM and the C-SKL states. Fig.~\hyperlink{anchor:Ecomp}{\ref{fig:E_comp}} compares the zero-temperature energies of the related topological phases, including the C-SKL, M-AM, and incommensurate skyrmion lattice (IC-SKL), as a function of varying strengths of easy-plane anisotropy in two representative magnetic field of 0.9 and 1.7~T. The calculations were performed with the optimization function in Sunny.jl \cite{dahlbomSunnyjl2025} on the original breathing kagome lattice using the nonlinear generalized minimal residual (NGMRES) algorithm. Compared to the conjugate gradient algorithm, the NGMRES algorithm is more stable in finding the optimal spin configuration while preserving the global topology of the spin textures. Starting from an initial state, the spin configuration was incrementally optimized at each magnetic field step of 0.005~T. This step-wise relaxation ensures that the system settles into the local energy minimum corresponding to a specific field response without disrupting the magnetic structure. As compared in Fig.~\hyperlink{anchor:Ecomp}{\ref{fig:E_comp}}, the energy gained through the SIA is not sufficient to induce a transition from the IC-SKL to the C-SKL in the considered field regime. As a contrast, in a relatively low field of 0.9~T, the energy of a M-AM lattice is already lower than that of the IC-SKL even with a relatively weak SIA of 0.001 meV and the relative stability of the M-AM lattice keeps increasing with the strength of SIA. In a relatively high field of 1.7 T, the IC-SKL becomes more stable for the fitted SIA strength of 0.009 meV, which explains the commensurate-incommensurate shift in the magnetic propagation vector observed in experiments~\cite{hirschbergerLatticecommensurate2024s}. Therefore, our microscopic modeling establishes the commensurate phase in Gd$_3$Ru$_4$Al$_{12}$ to be of a M-AM state rather than the previously proposed C-SKL state.

\section{Stability of the transverse conical phase}
\begin{figure*}[t]
    \centering
\renewcommand{\thefigure}{S\arabic{figure}}
    \includegraphics[width=0.6\linewidth]{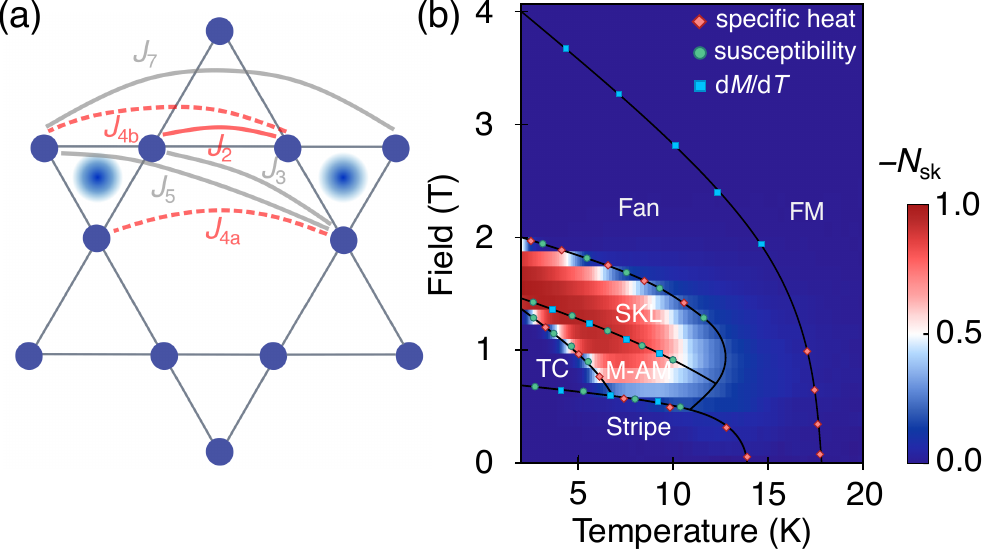}
    \hypertarget{anchor:TC}{}
    \caption{(a) A schematic illustrating the equivalence between the nearest-neighbor exchange interaction, $J_2^T$, on the effective triangular lattice and the third-neighbor interactions, $J_{3a}$ and $J_{3b}$, on the pristine lattice. (b) The temperature-field (T-H) phase diagram of the skyrmion number ($N_{sk}$) simulated on an 11$\times $11$\times $2 supercell of the original breathing kagome lattice. }
    \label{fig:TC}
\end{figure*}

Considering that the previous experiments revealed a field-induced non-topological transverse conical (TC) phase at low temperatures~\cite{hirschbergerSkyrmion2019s}, here we explore how to further stabilize the TC phase starting from our microscopic model. Firstly, through local structural optimization on the original breathing kagome lattice, we found that in a field of 1.0~T, the energy of the TC phase is only slightly higher than that of the M-AM phase by $\sim0.02$~meV per spin, which suggests that the TC phase may be stabilized as the ground state by fine tuning the coupling parameters. As explained in the main text, introducing additional exchange interactions among the trimers that are already coupled in the $J_{126}$-$J_{c123}$-$J_{n12}$ model does not impact the goodness-of-fit. As an illustration, Fig.~\hyperlink{anchor:TC}{\ref{fig:TC}(a)} plots all the couplings among the first-neighboring trimers on the original breathing kagome lattice: Besides the $J_2$ interactions that are considered in the $J_{126}$-$J_\mathrm{c123}$-$J_\mathrm{n12}$ model, there are also the $J_3$, $J_{4a}$, $J_{4b}$, $J_5$, and $J_7$ interactions that couple the same first-neighboring trimers, while adding these extra interactions into the model does not improve the fits. Based on this observation, we introduced counteracting  $J_\mathrm{4a}$ = 0.03 and $J_\mathrm{4b}$ = -0.015~meV into the model, and found that the TC phase become more stable than the M-AM phase by 0.003~meV per spin.

\begin{figure*}[b]
    \centering
\renewcommand{\thefigure}{S\arabic{figure}}
    \includegraphics[width=0.8\linewidth]{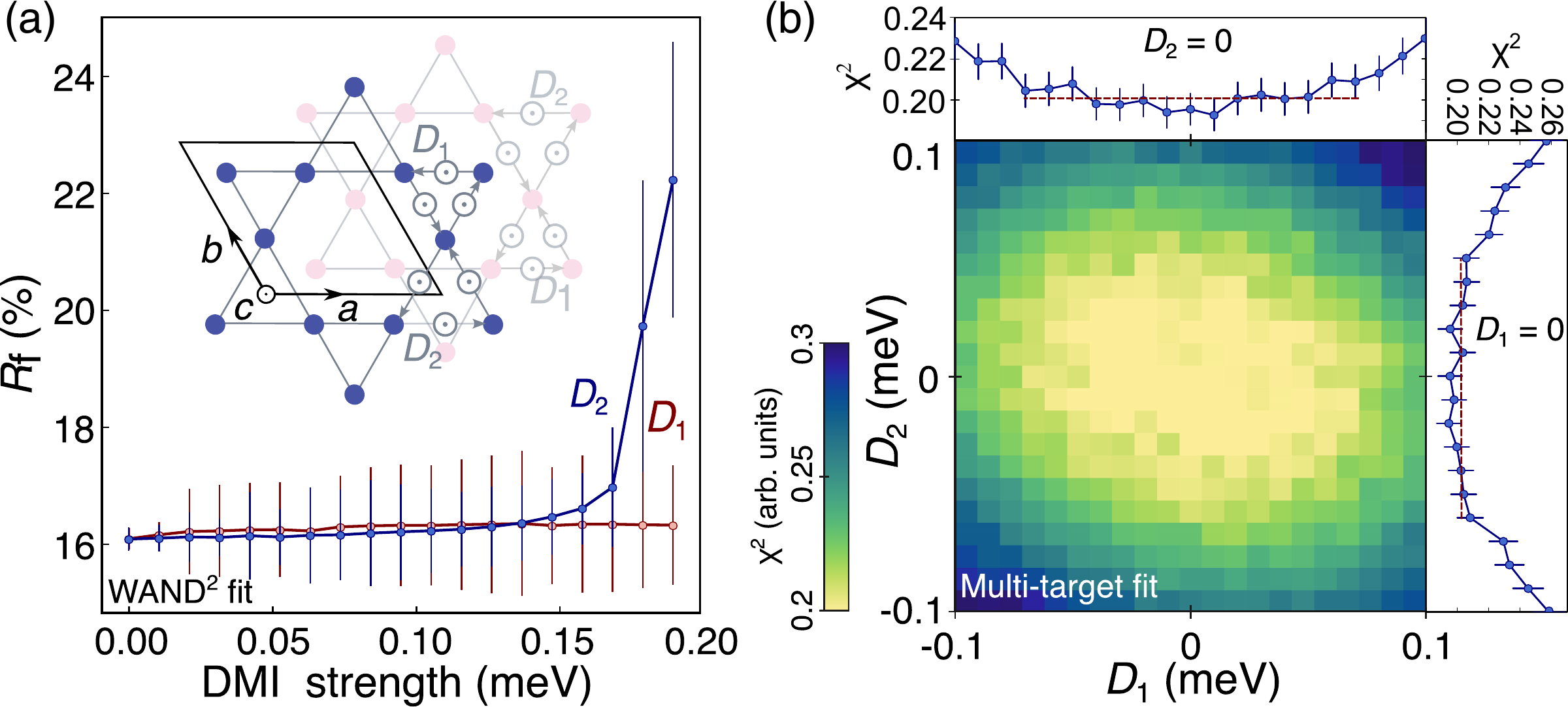}
    \hypertarget{anchor:DM_insensitivity}{}
    \caption{Impacts of the DM interactions on the fits. (a) The goodness-of-fit parameter, $R_\mathrm{f}$, between the experimental and calculated neutron diffraction intensity of the magnetic Bragg peaks as a function of $D_1$ and $D_2$. Inset describes the directions of the DM vectors over the two neighboring layers of breathing kagome lattices. (b) The goodness-of-fit parameter, $\chi^2$, as a function of $D_1$ and $D_2$. The right and the top sub-panel are slices at $D_1 =0$ and $D_2=0$~meV, respectively. The red dashed lines indicate the range of $|D_1|<0.05$~meV and $|D_2|<0.07$~meV, where the variation of $\chi^2$ is lower than its standard deviation.}
    \label{fig:DM_insensitivity}
\end{figure*}

The enhanced stability of the TC phase can be verified through our Monte Carlo calculations, and the results are summarized in Figs.~\hyperlink{anchor:TC}{\ref{fig:TC}(b)}. Monte Carlo simulations were performed on an 11$\times$11$\times$2 supercell of the original lattice under the parallel tempering scheme. It is observed that after introducing $J_\mathrm{4a}$ and $J_\mathrm{4b}$, a TC phase appears at 2~K in a field range of 0.8 and 1.4~T. This observation suggest that the discrepancy in the experimental and calculated TC phase may arise from additional terms not included in our model, such as redundant spin interactions among the trimers or longer range exchange interactions.

\begin{figure*}[b]
    \centering
\renewcommand{\thefigure}{S\arabic{figure}}
    \includegraphics[width=0.75\linewidth]{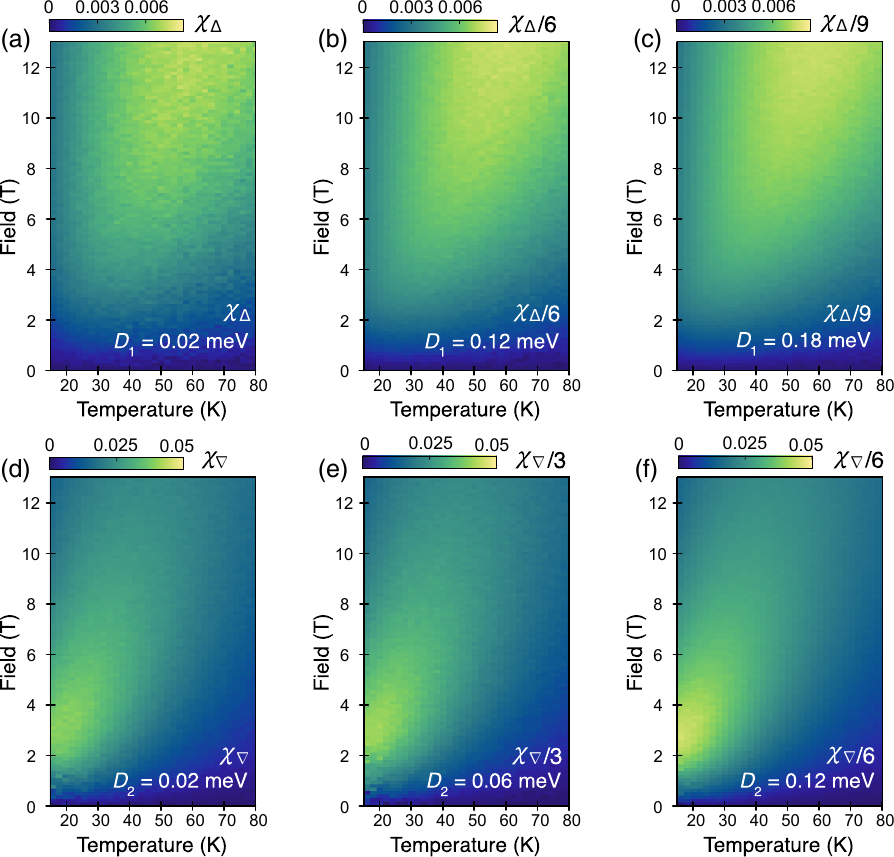}
    \hypertarget{anchor:DM_chirality}{}
    \caption{Impacts of the DM interactions on chiral fluctuations. Comparison of the chiral fluctuations with different DM interactions: (a) $D_1 = 0.02$~meV, (b) $D_1 = 0.12$~meV, (c) $D_1 = 0.18$~meV, (d) $D_2 = 0.02$~meV, (e) $D_2 = 0.06$~meV, and (f) $D_2 = 0.12$~meV. Pseudocolor represents the calculated value of chirality. Chirality in panels~(b,c,e,f) are divided by a factor of 6, 9, 3, and 6, respectively. The simulations are calculated on original lattice.}
    \label{fig:chirality}
\end{figure*}

\section{Impacts of the Dzyaloshinskii-Moriya interactions on the fits}

As shown in the inset of Fig.~\hyperlink{anchor:DM_insensitivity}{\ref{fig:DM_insensitivity}(a)}, staggered Dzyaloshinskii-Moriya (DM) interactions are allowed in Gd$_3$Ru$_4$Al$_{12}$ despite its centrosymmetry. To confirm that the introduction of additional DM interactions with a strength of $\sim 0.02$~meV as considered in the main text do not impact the fitting of the experimental data, we first clarify the impacts of the DM interactions on the ground state. Fig.~\hyperlink{anchor:DM_insensitivity}{\ref{fig:DM_insensitivity}(a)} presents the goodness-of-fit factor, $R_\mathrm{f}$, as a function of the strengths of the DM interactions, including $D_1$ over the smaller triangles and $D_2$ over the larger triangles. Errors are estimated from 930 independent runs of energy minimization with random initial spin configurations. Due to the dominating strength of the $J_1$ interactions, the introduction of $D_1$ up to 0.20~meV do not impact the ground state. Over the larger triangles, the threshold of $D_2$ is estimated to be $\sim0.12$~meV, above which the $R_\mathrm{f}$ starts to deviate from the optimal value.

Next, we study the impacts of the DM interactions on the multi-target fits. Additional DM interactions, including $D_1$ and $D_2$, are introduced to the $J_{126}$-$J_{\rm{c123}}$-$J_{\rm{n12}}$ model on the original breathing kagome lattice with all parameters fixed at the optimal values in the absence of DM interactions. Therefore, our calculations provide the upper limit for  $\chi^2$ at each values of $D_1$ and $D_2$. Figure~\hyperlink{anchor:DM_insensitivity}{\ref{fig:DM_insensitivity}(b)} presents the calculated $\chi^2$ as a function of $D_1$ and $D_2$. Slices shown on the right and top sub-panels are calculated at $D_1=0$ and $D_2=0$, respectively. Due to the high computation cost, we estimate the errors by assuming the same ratio between the errors and values of $\chi^2$ for each combination of $D_1$ and $D_2$, which contains altogether 400 samplings. For each sampling, we calculate the $\chi^2$ twice, and their differences are found to follow a Gaussian distribution, from which the FWHM provides an estimate of the standard deviations. As shown in Fig.~\hyperlink{anchor:DM_insensitivity}{\ref{fig:DM_insensitivity}(b)}, the variation of $\chi^2$ is lower than its standard deviation in the range of $|D_1|<0.05$~meV and $|D_2|<0.07$~meV. Therefore, we conclude that the introduction of additional DM interactions with a strength of 0.02~meV does not impact the fits of the experimental data.
\section{Further calculations of the chiral fluctuations in the spiral spin-liquid phase}
In the main text, the thermally activated scalar spin chirality in the spiral spin-liquid phase is calculated by

\renewcommand{\theequation}{S\arabic{equation}} 
\begin{align}
\chi_\Delta &= \left<\frac{1}{N_\Delta}\left| \sum_{i,j,k \in \Delta}\mathbf{S}_i\cdot (\mathbf{S}_j \times \mathbf{S}_k)\right|\right>\textrm{,}\\
\chi_\nabla &= \left<\frac{1}{N_\nabla}\left|\sum_{i,j,k \in \nabla}\mathbf{S}_i\cdot (\mathbf{S}_j \times \mathbf{S}_k)\right|\right>\textrm{,}
\end{align}
where $\left<...\right>$ represents the thermal average. $\Delta$ and $\nabla$ denote the smaller and larger triangles, respectively. The summations run over all triangles belonging to the set $\Delta$ and $\nabla$, with $N_\Delta$ and $N_\nabla$ denoting their total numbers, respectively. The quantity $\mathbf{S}_i\cdot  (\mathbf{S}_j\times \mathbf{S}_k)$ is the local scalar spin chirality associated with the specific triangle defined on three vertices at $i$, $j$, and $k$, which supplies a fictitious magnetic field over the conduction electrons and thus induces an anomalous Hall effect~\cite{nagaosa_anomalous_2010s}.
\begin{figure*}[b]
    \centering
\renewcommand{\thefigure}{S\arabic{figure}}
    \includegraphics[width=0.4\linewidth]{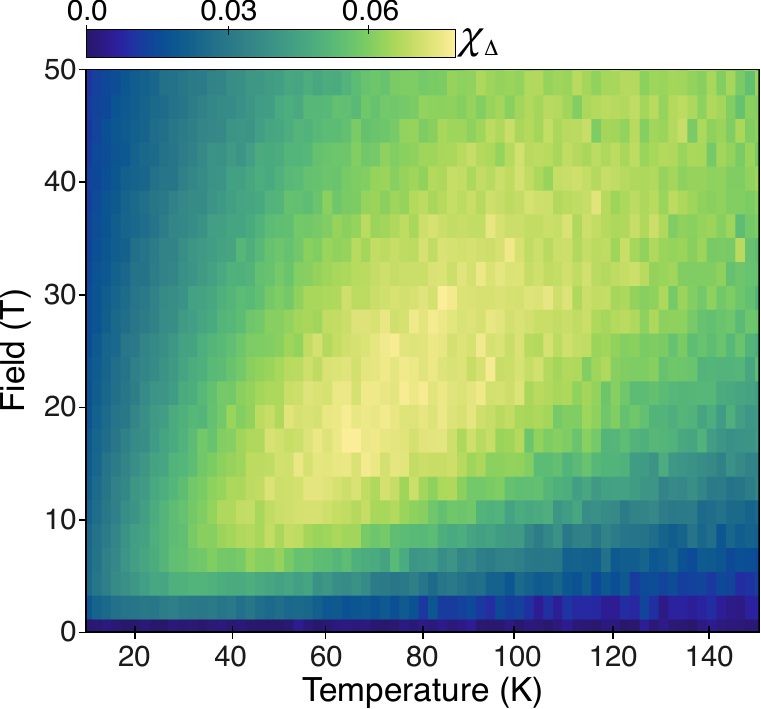}
    \hypertarget{anchor:DM_chirality}{}
    \caption{Calculated $\chi_\Delta$ in an extend region. Scalar spin-chirality, $\chi_\Delta$, over the smaller triangles calculated for $D_1 = 0.02$~meV in an extended region of $T = [10, 150]$~K and $H = [0, 50]$~T.} 
    \label{fig:chirality1}
\end{figure*}

Figure~\hyperlink{anchor:chirality}{\ref{fig:chirality}} presents further comparisons of the calculated scalar spin chirality in the presence of different strengths of $D_1$ and $D_2$, which reveal that the field and temperature variations of the chiral fluctuations are mainly determined by the types of DM interactions rather than their exact strength. For example, as shown in Figs.~\hyperlink{anchor:chirality}{\ref{fig:chirality}(a-c)}, increasing $D_1$ from 0.02 to 0.18~meV yields a chirality distribution that remains largely unchanged, with the chiral fluctuations being continuously enhanced at increased temperatures and field strengths. While for $D_2$, as compared in Figs.\hyperlink{anchor:chirality}~\hyperlink{anchor:chirality}{\ref{fig:chirality}(d-f)}, the maximum of the chiral fluctuation stays at $\sim3$~T at low temperatures independent of the strengths of $D_2$, which is similar to the experimental observations \cite{kolincioKagome2023s}. As presented in Fig.~\hyperlink{anchor:chirality1}{\ref{fig:chirality1}}, the scalar spin chirality over the smaller triangles, $\chi_\Delta$, is also calculated in an extended range of magnetic fields and temperatures. The maximum of $\chi_\Delta$ is observed in a broad area centered around $H$ $\sim$ 20~T and $T$ $\sim$ 80~K. This comparison confirms the critical role of scalar spin-chirality over the $J_2$ bonds in explaining the experimentally observed anamlous Hall effect~\cite{kolincioKagome2023s}.

\end{document}